\documentclass[showkeys,superscriptaddress,floatfix,prd,10pt,aps]{revtex4-2}
\usepackage{graphicx,epstopdf}
\usepackage[caption=false]{subfig}
\usepackage{appendix}
\usepackage[T1]{fontenc}
\usepackage{lmodern}
\usepackage[dvipsnames,x11names]{xcolor}
\usepackage[colorlinks=true,linkcolor=Magenta!80!black,citecolor=Magenta!70!black,urlcolor=Magenta!80!black]{hyperref}
\usepackage[sort&compress]{natbib}
\usepackage{lipsum}
\usepackage{morefloats}
\usepackage[pdf]{pstricks}
\usepackage{amsmath}
\usepackage{amssymb}
\usepackage{amsfonts}
\usepackage{rotating}
\usepackage{cancel}
\usepackage{mathtools}
\usepackage{bbm}
\usepackage{dsfont}
\usepackage{bbold}
\usepackage{multirow}
\usepackage{ulem}
\usepackage{physics}
\usepackage{orcidlink}
\usepackage{colortbl}
\usepackage{slashed}

\usepackage{booktabs}
 \usepackage{xcolor}
 \usepackage{colortbl}
 \usepackage{array}

\def\p_1slash{p1\!\!\!\slash }
\def\p_2slash{p2\!\!\!\slash }
\def\qslash{q\!\!\!\slash }

\def\eslash{\varepsilon\!\!\!\slash }

\definecolor{headerblue}{RGB}{28, 66, 110}
\definecolor{rowgray}{RGB}{245, 247, 250}

\begin{document}

\title{Charting doubly strange hidden-charm pentaquarks: An electromagnetic mapping of spin-$\frac{1}{2}$ and $\frac{3}{2}$ states}

\author{Ula\c{s}~\"{O}zdem\orcidlink{0000-0002-1907-2894}}%
\email[]{ulasozdem@aydin.edu.tr}
\affiliation{ Health Services Vocational School of Higher Education, Istanbul Aydin University, Sefakoy-Kucukcekmece, 34295 Istanbul, T\"{u}rkiye}

\begin{abstract}
We calculate the magnetic dipole moments of doubly strange hidden-charm pentaquark states with spin-parity assignments $J^P = \frac{1}{2}^-$ and $J^P = \frac{3}{2}^-$ using QCD light-cone sum rules---presenting the first systematic QCD light-cone sum rule investigation of the electromagnetic multipole structure in the $S = -2$ sector. To assess the model dependence of the predictions, we employ a set of independent interpolating currents constructed in diquark-diquark-antiquark form, which probe different assumptions about the internal color-spin correlations. For the spin-$\frac{3}{2}$ states, we also compute the electric quadrupole and magnetic octupole moments as complementary observables. The magnetic dipole moments exhibit a considerable spread across different currents: they range from $-2.15\,\mu_N$ to $5.74\,\mu_N$ for spin-$\frac{1}{2}$ pentaquarks and from $-4.25\,\mu_N$ to $-0.43\,\mu_N$ for their spin-$\frac{3}{2}$ counterparts. This variation reflects the sensitivity of magnetic moments to the detailed internal wave function. A quark-level decomposition reveals that the charm quark provides the dominant contribution in most configurations, while strange quarks play a decisive role only in currents that favor axial-vector diquark structures. For the spin-$\frac{3}{2}$ states, the electric quadrupole moments lie between $-2.01\times10^{-2}$~fm$^2$ and $5.55\times10^{-2}$~fm$^2$, and the magnetic octupole moments are typically an order of magnitude smaller. The significant current dependence of the magnetic dipole moments provides a quantitative measure of the theoretical uncertainty arising from the choice of interpolating operator. The pronounced isospin sensitivity of $J^3_\mu(x)$ across all three multipole moments is shown to arise from its axial-vector diquark structure, which isolates the light quark from spin averaging and allows the charge asymmetry $e_u/e_d = -2$ to propagate directly into the electromagnetic moments; the ratio $\mu_u/\mu_d = -2.00$ confirms this mechanism exactly.
Our predictions offer concrete benchmarks for future experimental measurements and lattice QCD calculations, and they may help discriminate among competing structural models for doubly strange hidden-charm pentaquarks.
\end{abstract}

\maketitle

\section{Introduction}\label{sec:introduction}

The possibility of hadronic states beyond the conventional quark-antiquark mesons and three-quark baryons has been a subject of theoretical speculation for decades. Experimental confirmation of such exotic configurations began in 2003 with the Belle Collaboration's discovery of the $X(3872)$~\cite{Belle:2003nnu}, whose properties defy explanation within the standard meson picture, establishing it as the first compelling tetraquark candidate. Since then, an increasingly rich spectrum of exotic hadrons has been uncovered, challenging and extending our understanding of  nonperturbative dynamics of QCD. These discoveries have triggered intensive theoretical efforts to interpret their nature, with proposals ranging from compact multiquark states and hadronic molecules to kinematic threshold effects~\cite{Esposito:2014rxa, Esposito:2016noz, Olsen:2017bmm, Lebed:2016hpi, Nielsen:2009uh, Brambilla:2019esw, Agaev:2020zad, Chen:2016qju, Ali:2017jda, Guo:2017jvc, Liu:2019zoy, Yang:2020atz, Dong:2021juy, Dong:2021bvy, Chen:2022asf, Meng:2022ozq}. 

A landmark achievement in exotic hadron spectroscopy came in 2015, when the LHCb Collaboration reported two pentaquark states, $P_{\psi}^{N}(4380)$ and $P_{\psi}^{N}(4450)$, in the $J/\psi p$ spectrum from $\Lambda^0_b \to J/\psi p K^-$ decays~\cite{LHCb:2015yax}. Subsequent analyses with increased statistics refined this picture, revealing the narrow $P_{\psi}^{N}(4312)$ and resolving the $P_{\psi}^{N}(4450)$ into two distinct peaks, $P_{\psi}^{N}(4440)$ and $P_{\psi}^{N}(4457)$~\cite{LHCb:2019kea}. With minimal quark content $uudc\bar{c}$, these states opened a new chapter in the study of hidden-charm pentaquarks. The search was soon extended to strange sectors. In 2020, LHCb presented evidence for $P^{\Lambda}_{\psi s}(4459)$ in the $J/\psi\Lambda$ channel from $\Xi^-_b \to J/\psi\Lambda K^-$ decays~\cite{LHCb:2020jpq}, followed in 2022 by the observation of $P^{\Lambda}_{\psi s}(4338)$ in the same final state~\cite{LHCb:2022ogu}. These candidates, with quark content $udsc\bar{c}$, represent the first hints of hidden-charm pentaquarks carrying strangeness. Very recently, the Belle Collaboration has reported a $P^{\Lambda}_{\psi s}$-like signal at $(4471.7 \pm 4.8 \pm 0.6)$ MeV with a width of $(21.9 \pm 13.1 \pm 2.7)$ MeV, reaching $3.3\sigma$ significance including systematic uncertainties~\cite{Belle:2025pey}. Interestingly, the data do not exclude the possibility that this signal arises from two closely spaced resonances, a scenario discussed in concurrent theoretical work~\cite{Clymton:2025hez}.  However, the interpretation of these states remains contentious. Their spins and parities are experimentally unconstrained, and fundamental questions about their internal organization persist: are they compact multiquark clusters, weakly bound hadronic molecules, or perhaps threshold cusps enhanced by final-state interactions? Discriminating among these scenarios requires observables that probe aspects of hadron structure complementary to mass and decay width. 

Electromagnetic moments, especially the magnetic dipole moment, serve precisely this purpose. As leading-order responses to an external magnetic field, they encode information about the spatial distribution of circulating currents and the alignment of quark spins within the hadron. For a composite system, the magnetic moment depends not only on the charges and spins of the constituents but also on their relative spatial wave function—making it a sensitive probe of internal geometry and correlation patterns. In the context of exotic multiquark states, different structural models (e.g., tightly bound diquark-triquark systems versus extended molecular configurations) predict characteristically different magnetic moments, owing to distinct arrangements of light and heavy quarks and to variations in the internal orbital angular momentum. Thus, theoretical predictions for these moments provide valuable benchmarks that can, in conjunction with future measurements, help clarify the underlying nature of the observed resonances.

Motivated by these considerations, we undertake a systematic investigation of the magnetic dipole, electric quadrupole and magnetic octupole moments of doubly strange hidden-charm pentaquarks ($S = -2$) within the QCD light-cone sum rule framework~\cite{Chernyak:1990ag, Braun:1988qv, Balitsky:1989ry}. We employ a set of interpolating currents built from diquark-diquark-antiquark operators, targeting states with $J^P = 1/2^-$ and $3/2^-$. Since different choices of Dirac and color structure in these currents encode distinct hypotheses about the internal spin–color correlations, studying how the predicted moments vary across currents offers insight into the structural sensitivity of electromagnetic observables. Our analysis provides predictions for magnetic dipole moments and, where applicable, for electric quadrupole and magnetic octupole moments, thereby supplying quantitative benchmarks that may aid future experimental searches and help discriminate among competing structural pictures. 
The electromagnetic properties of exotic hadrons have been studied extensively in recent years, with particular attention to hidden-charm and hidden-bottom pentaquarks using various theoretical approaches~\cite{Ozdem:2025jda, Wang:2016dzu, Ozdem:2018qeh, Ortiz-Pacheco:2018ccl, Xu:2020flp, Ozdem:2021btf, Ozdem:2021ugy, Li:2021ryu,   Ozdem:2023htj, Wang:2023iox, Ozdem:2022kei, Gao:2021hmv, Ozdem:2024rch, Guo:2023fih, Ozdem:2022iqk, Wang:2022nqs, Wang:2022tib, Ozdem:2024jty, Li:2024wxr, Li:2024jlq, Ozdem:2024yel, Ozdem:2024rqx, Mutuk:2024ltc, Mutuk:2024jxf, Ozdem:2024usw, Ozdem:2025fks, Zhu:2025abk, Ozdem:2025ncd, Ozdem:2025ion, Mutuk:2024ach}. The present work extends this program to the doubly strange sector—which has received comparatively less attention—and explicitly addresses the model dependence associated with the choice of interpolating current, an aspect essential for robust uncertainty estimation and for understanding which structural features electromagnetic moments actually constrain.

The paper is organized as follows. Section~\ref{sec:formalism} outlines the  QCD light-cone sum rule framework, including the construction of correlation functions and interpolating currents. The hadronic and QCD representations of the sum rules are derived, leading to master formulas for the multipole moments. Section~\ref{sec:numerical} presents the numerical analysis, discussing input parameters, working regions for Borel mass and continuum threshold, and the resulting predictions for electromagnetic moments. A quark-level decomposition of the moments is provided to illuminate the underlying dynamics. Section~\ref{sec:conclusions} summarizes our findings and discusses their implications for future experimental and theoretical studies of exotic hadrons.

\begin{widetext}
 
\section{Construction of the QCD light-cone sum rules}\label{sec:formalism}

The calculation of electromagnetic properties within the  QCD light-cone sum rule (LCSR) approach begins with the definition of an appropriate vacuum correlation function in the background of a soft photon field. This function serves as the central object that connects the hadronic matrix elements of interest to the underlying quark-gluon dynamics. For the doubly strange hidden-charm pentaquark states, we introduce two correlation functions corresponding to the spin-1/2 and spin-3/2 sectors:
\begin{align}
\label{eq:correlator_12}
\Pi(p, q) &= i \int d^4x\, e^{ip\cdot x} \, 
\bigl\langle 0 \bigl| \mathrm{T}\bigl\{ J(x) \,
\bar{J}(0) \bigr\} \bigr| 0 \bigr\rangle_{\gamma}, \\[6pt]
\label{eq:correlator_32}
\Pi_{\mu\nu}(p, q) &= i \int d^4x\, e^{ip\cdot x} \, 
\bigl\langle 0 \bigl| \mathrm{T}\bigl\{ J_{\mu}(x) \,
\bar{J}_{\nu}(0) \bigr\} \bigr| 0 \bigr\rangle_{\gamma}.
\end{align}
Here, \(p\) is the four-momentum of the incoming pentaquark state, \(q\) denotes the four-momentum carried by the external photon with polarization vector \(\varepsilon_{\mu}(q)\), and the subscript \(\gamma\) indicates that the vacuum expectation value is taken in the presence of the electromagnetic background. The local composite operators \(J(x)\) and \(J_{\mu}(x)\) are interpolating currents constructed to possess the same quantum numbers as the spin-1/2 and spin-3/2 doubly strange hidden-charm pentaquarks, respectively. These currents act as sources that can create the states of interest from the QCD vacuum.

\subsection{Interpolating currents for doubly strange hidden-charm pentaquarks}

The choice of interpolating current represents a specific ansatz for the internal wave function of the hadron. To systematically explore the model dependence of our predictions and to cover a range of plausible internal configurations, we employ multiple independent currents for each spin-parity assignment. These currents are built from diquark-type substructures—with scalar and axial-vector configurations being phenomenologically favored~\cite{Wang:2010sh, Kleiv:2013dta}—combined with an explicit anti-charm quark to form overall color-singlet objects.  
The diquark-based interpretation of hidden-charm pentaquarks has been extensively pursued in the literature since the original LHCb observation, including the diquark-diquark-antiquark analyses~\cite{Maiani:2015vwa, Lebed:2015tna} and the related diquark-triquark picture~\cite{Zhu:2015bba}, which together established the compact multiquark framework adopted in the present work.

For the \(J^{P} = \frac{1}{2}^{-}\) pentaquark sectors, we consider the following four independent interpolating currents:
\begin{eqnarray}
\label{eq:current_1_12}
J_{1}(x) &=& \varepsilon_{abc}\,\varepsilon_{ade}\,\varepsilon_{bfg}\;
\Bigl[ q^{T}_{d}(x) C \gamma_{5} s_{e}(x) \Bigr]
\Bigl[ s^{T}_{f}(x) C \gamma_{5} c_{g}(x) \Bigr]
C \bar{c}^{T}_{c}(x), \\[6pt]
\label{eq:current_2_12}
J_{2}(x) &=& \varepsilon_{abc}\,\varepsilon_{ade}\,\varepsilon_{bfg}\;
\Bigl[ q^{T}_{d}(x) C \gamma_{5} s_{e}(x) \Bigr]
\Bigl[ s^{T}_{f}(x) C \gamma_{\alpha} c_{g}(x) \Bigr]
\gamma_{5}\gamma^{\alpha} C \bar{c}^{T}_{c}(x), \\[6pt]
\label{eq:current_3_12}
J_{3}(x) &=& \varepsilon_{abc}\,\varepsilon_{ade}\,\varepsilon_{bfg}\;
\Bigl[ s^{T}_{d}(x) C \gamma_{\alpha} s_{e}(x) \Bigr]
\Bigl[ q^{T}_{f}(x) C \gamma_{5} c_{g}(x) \Bigr]
\gamma_{5}\gamma^{\alpha} C \bar{c}^{T}_{c}(x), \\[6pt]
\label{eq:current_4_12}
J_{4}(x) &=& \varepsilon_{abc}\,\varepsilon_{ade}\,\varepsilon_{bfg}\;
\Bigl[ s^{T}_{d}(x) C \gamma_{\alpha} s_{e}(x) \Bigr]
\Bigl[ q^{T}_{f}(x) C \gamma^{\alpha} c_{g}(x) \Bigr]
C \bar{c}^{T}_{c}(x).
\end{eqnarray}

For the \(J^{P} = \frac{3}{2}^{-}\) sectors, the corresponding set of currents with a Lorentz vector index is:
\begin{eqnarray}
\label{eq:current_1_32}
J^{1}_{\mu}(x) &=& \varepsilon_{abc}\,\varepsilon_{ade}\,\varepsilon_{bfg}\;
\Bigl[ q^{T}_{d}(x) C \gamma_{5} s_{e}(x) \Bigr]
\Bigl[ s^{T}_{f}(x) C \gamma_{\mu} c_{g}(x) \Bigr]
C \bar{c}^{T}_{c}(x), \\[6pt]
\label{eq:current_2_32}
J^{2}_{\mu}(x) &=& \varepsilon_{abc}\,\varepsilon_{ade}\,\varepsilon_{bfg}\;
\Bigl[ s^{T}_{d}(x) C \gamma_{\mu} s_{e}(x) \Bigr]
\Bigl[ q^{T}_{f}(x) C \gamma_{5} c_{g}(x) \Bigr]
C \bar{c}^{T}_{c}(x), \\[6pt]
\label{eq:current_3_32}
J^{3}_{\mu}(x) &=& \varepsilon_{abc}\,\varepsilon_{ade}\,\varepsilon_{bfg}\;
\Bigl[ s^{T}_{d}(x) C \gamma_{\mu} s_{e}(x) \Bigr]
\Bigl[ q^{T}_{f}(x) C \gamma_{\alpha} c_{g}(x) \Bigr]
\gamma_{5}\gamma^{\alpha} C \bar{c}^{T}_{c}(x), \\[6pt]
\label{eq:current_4_32}
J^{4}_{\mu}(x) &=& \varepsilon_{abc}\,\varepsilon_{ade}\,\varepsilon_{bfg}\;
\Bigl[ s^{T}_{d}(x) C \gamma_{\alpha} s_{e}(x) \Bigr]
\Bigl[ q^{T}_{f}(x) C \gamma_{\mu} c_{g}(x) \Bigr]
\gamma_{5}\gamma^{\alpha} C \bar{c}^{T}_{c}(x).
\end{eqnarray}

In these expressions, the indices \(a, b, c, d, e, f, g\) are color indices; \(q\) denotes a light \(u\) or \(d\) quark; \(C = i\gamma^{2}\gamma^{0}\) is the charge-conjugation matrix; the superscript \(T\) indicates transpose in Dirac space. The Levi-Civita symbols $\varepsilon$ enforce full antisymmetrization in color space: each diquark pair forms a color antitriplet, and the overall five-quark operator is a color singlet. The Dirac structures determine the spin content: $C\gamma_5$ creates scalar (spin-0) diquarks, while $C\gamma_\mu$ creates axial-vector (spin-1) diquarks. The additional matrices ($\gamma_5\gamma^\alpha$ or unity) coupling these clusters to the anti-charm quark ensure the desired total spin-parity $J^P = \frac{1}{2}^-$ or $\frac{3}{2}^-$. The variation in Dirac structure among the currents—changing the gamma matrices coupling the diquark pairs and the coupling to the anti-charm quark—allows us to scan different possibilities for the internal spin–orbital wave function of the pentaquark. The resulting spread in the predicted electromagnetic moments across this set of currents serves a dual purpose: it provides a quantitative measure of the theoretical uncertainty arising from our imperfect knowledge of the true wave function, and it reveals how sensitively electromagnetic observables probe specific aspects of the quark–gluon substructure, such as diquark spin correlations and relative orbital angular momentum. 
The four currents employed here span the complete set of independent diquark--diquark--antiquark operators constructed from scalar ($C\gamma_5$) and axial-vector ($C\gamma_\mu$) diquark substructures; operators involving higher Dirac structures such as $C\gamma_5\gamma_\mu$ or $C\sigma_{\mu\nu}$ are phenomenologically disfavored~\cite{Wang:2010sh, Kleiv:2013dta} and are not considered here.
 
\subsection{Phenomenological representation of the correlation functions}
\label{subsec:hadronic}

The phenomenological (hadronic) side of the correlation functions is obtained by inserting a complete set of intermediate states with the same quantum numbers as the interpolating currents. For the spin-1/2 and spin-3/2 pentaquark channels, the resulting spectral representations read:
\begin{align}
\label{eq:phen_12}
\Pi^{\mathrm{phen}}(p,q) &= 
\frac{\langle 0 | J(0) | P^{\Lambda}_{\psi ss}(p,s)\rangle}
{p^{2} - m^{2}} \,
\langle P^{\Lambda}_{\psi ss}(p,s) | P^{\Lambda}_{\psi ss}(p+q,s)\rangle_{\gamma} \,
\frac{\langle P^{\Lambda}_{\psi ss}(p+q,s) | \bar{J}(0) | 0 \rangle}
{(p+q)^{2} - m^{2}} + \cdots , \\[8pt]
\label{eq:phen_32}
\Pi^{\mathrm{phen}}_{\mu\nu}(p,q) &= 
\frac{\langle 0 | J_{\mu}(0) | P^{\ast\Lambda}_{\psi ss}(p,s)\rangle}
{p^{2} - { m^\ast}^{2}} \,
\langle P^{\ast\Lambda}_{\psi ss}(p,s) | P^{\ast\Lambda}_{\psi ss}(p+q,s)\rangle_{\gamma} \,
\frac{\langle P^{\ast\Lambda}_{\psi ss}(p+q,s) | \bar{J}_{\nu}(0) | 0 \rangle}
{(p+q)^{2} - { m^\ast}^{2}} + \cdots ,
\end{align}
where the quantities $m$ and ${m^\ast}$ correspond to the masses of the spin-$1/2$ and spin-$3/2$ states, respectively. The ellipses denote contributions from higher resonances and the continuum of states, which will be treated using the quark-hadron duality prescription. The necessary matrix elements are parametrized in terms of hadronic observables as follows. The coupling of the interpolating currents to the physical pentaquark states is defined through the pole residues \(\lambda_{P}\): 
\begin{align}
\label{eq:pole_12}
\langle 0 | J(0) | P^{\Lambda}_{\psi ss}(p,s)\rangle &= \lambda_{P^{\Lambda}_{\psi ss}} \, \gamma_{5} \, u(p,s), \\[4pt]
\label{eq:pole_32}
\langle 0 | J_{\mu}(0) | P^{\ast\Lambda}_{\psi ss}(p,s)\rangle &= \lambda_{P^{\ast\Lambda}_{\psi ss}} \, u_{\mu}(p,s),
\end{align}
where \(u(p,s)\) and \(u_{\mu}(p,s)\) are the Dirac and Rarita--Schwinger spinors for spin-1/2 and spin-3/2 particles, respectively, satisfying the standard normalization conditions. The electromagnetic vertex of the pentaquark in the background photon field is parametrized by a set of gauge-invariant form factors. For the spin-1/2 transition, we adopt the conventional decomposition \cite{Leinweber:1990dv}:
\begin{align}
\label{eq:vertex_12}
\langle P^{\Lambda}_{\psi ss}(p,s) | P^{\Lambda}_{\psi ss}(p+q,s)\rangle_{\gamma} &= 
\varepsilon^{\mu} \, \bar{u}(p,s) \Bigl[ f_{1}(q^{2}) \gamma_{\mu} 
+ \frac{f_{2}(q^{2})}{2 m} \, i\sigma_{\mu\nu} q^{\nu} \Bigr] u(p+q,s),
\end{align}
while for the spin-3/2 transition, the most general parity-conserving vertex can be written as \cite{Nozawa:1990gt,Pascalutsa:2006up,Ramalho:2009vc}:
\begin{eqnarray}
\langle P^{\ast \Lambda}_{\psi ss}(p_2,s)\mid P^{\ast \Lambda}_{\psi ss}(p_1,s)\rangle_F &=&-e \,\bar
u_{\mu}(p_2,s)\Gamma_{\mu\nu} u_{\nu}(p_1,s),\label{matelpar}
\end{eqnarray}
with
\begin{align}
\label{eq:vertex_32}
 \Gamma_{\mu\nu} &= F_{1}(q^2)g_{\mu\nu}\eslash-
\frac{1}{2{m^\ast}}
\bigg[F_{2}(q^2)g_{\mu\nu} \eslash\qslash+F_{4}(q^2)\frac{q_{\mu}q_{\nu} \eslash\qslash}{(2m^\ast)^2}\bigg]
+
\frac{F_{3}(q^2)}{(2{m^\ast})^2}q_{\mu}q_{\nu}\eslash .
\end{align}
Here, \(\varepsilon^{\mu}\) is the photon polarization vector, \(e\) is the elementary charge, and \(f_{i}(q^{2})\), \(F_{i}(q^{2})\) are dimensionless electromagnetic form factors that encode the internal structure of the pentaquark. The magnetic dipole form factors, which are the primary focus of this work, are obtained from these vertex functions in the limit of a real photon (\(q^{2} \to 0\)). For the spin-1/2 case, the magnetic form factor is simply 
\[ F_{M}(q^{2}) = f_{1}(q^{2}) + f_{2}(q^{2}),\] 
so that the magnetic dipole moment  is given by
\begin{equation}
\label{eq:mu_12}
\mu_{P^{\Lambda}_{\psi ss}} = \frac{e}{2 m} \bigl[ f_{1}(0) + f_{2}(0) \bigr].
\end{equation}

For the spin-3/2 pentaquark, the magnetic dipole form factor is conventionally defined as \cite{Nozawa:1990gt,Pascalutsa:2006up}:  
\[G_{M}(q^{2}) = \bigl[ F_{1}(q^{2}) + F_{2}(q^{2}) \bigr]\Bigl( 1 + \frac{4}{5}\eta \Bigr) 
- \frac{2}{5} \bigl[ F_{3}(q^{2}) + F_{4}(q^{2}) \bigr]\eta (1 + \eta),\] 
where \(\eta = -q^{2}/(4 {m^\ast}^{2})\). In the static limit \(q^{2} \to 0\), this reduces to
\begin{equation}
\label{eq:mu_32}
\mu_{P^{\ast\Lambda}_{\psi ss}} = \frac{e}{2 {m^\ast}} \bigl[ F_{1}(0) + F_{2}(0) \bigr],
\end{equation}
which defines the magnetic dipole moment of the spin-3/2 state.

Substituting the parametrizations Eqs.~(\ref{eq:pole_12})--(\ref{eq:vertex_32}) into the spectral representations Eqs.~(\ref{eq:phen_12}) and (\ref{eq:phen_32}) and carrying out the necessary algebraic manipulations, we arrive at explicit expressions for the correlation functions. For the spin-1/2 channel, the piece containing the magnetic dipole form factors can be written as
\begin{align}
\label{eq:phen_explicit_12}
\Pi^{\mathrm{phen}}(p,q)  &= 
\frac{\lambda_{P^{\Lambda}_{\psi ss}}^{2}}
{[p^{2} - m^{2}][(p+q)^{2} -m^{2}]} 
\Bigl[ f_{1}(q^{2}) + f_{2}(q^{2}) \bigr] 
\slashed{p} \slashed{\varepsilon} \slashed{q} 
+ \cdots . \Bigr]
\end{align}

For the spin-3/2 case, the part of the correlation function that carries the magnetic dipole form factors takes the form
\begin{align}
\label{eq:phen_explicit_32}
\Pi^{\mathrm{phen}}_{\mu\nu}(p,q) &= 
\frac{\lambda_{P^{\ast\Lambda}_{\psi ss}}^{2}}
{[p^{2} - {m^\ast}^{2}][(p+q)^{2} - {m^\ast}^{2}]} 
\Bigl[ g_{\mu\nu} \slashed{p} \slashed{\varepsilon} \slashed{q} \, F_{1}(q^{2})
- {m^\ast} g_{\mu\nu} \slashed{\varepsilon} \slashed{q} \, F_{2}(q^{2}) 
- \frac{F_{3}(q^{2})}{4{m^\ast}} q_{\mu} q_{\nu} \slashed{\varepsilon} \slashed{q}\nonumber \\
& 
- \frac{F_{4}(q^{2})}{4 {m^\ast}^{3}} (\varepsilon \cdot p) q_{\mu} q_{\nu} \slashed{p} \slashed{q} + \cdots \Bigr].
\end{align}

The ellipses in both expressions denote other independent Lorentz structures that are not related to the magnetic dipole and other higher multipole moments. The coefficients of the structures shown above will be matched to the corresponding operator product expansion (OPE) results in the deep Euclidean region to derive the sum rules for the magnetic form factors \(f_{1}(0)+f_{2}(0)\) and \(F_{1}(0)+F_{2}(0)\), and hence for the magnetic dipole moments via Eqs.~(\ref{eq:mu_12}) and (\ref{eq:mu_32}).

\subsection{QCD representation of the correlation functions}
\label{subsec:qcd_representation}

The QCD representation of the correlation functions is derived by evaluating them in terms of fundamental quark and gluon degrees of freedom. This requires applying the OPE near the light-cone, which systematically incorporates both perturbative quark propagators and non-perturbative matrix elements parameterized through photon distribution amplitudes (DAs). To obtain the explicit form of the correlation function at the quark level, we apply Wick's theorem to contract the quark fields within the time-ordered product. As a concrete example, we illustrate this procedure for the interpolating currents \(J_1(x)\) and \(J_\mu^1(x)\) defined in Eqs.~(\ref{eq:current_1_12}) and (\ref{eq:current_1_32}). After carrying out all possible contractions, the correlation function for these currents factorizes into a product of quark propagators:
\begin{align}
\label{eq:qcd_explicit}
\Pi^{\mathrm{QCD}}_{1}(p,q) &= -i \, \varepsilon^{abc}\varepsilon^{a'b'c'}\varepsilon^{ade}\varepsilon^{a'd'e'}\varepsilon^{bfg} \varepsilon^{b'f'g'} 
\int d^4x \, e^{ip\cdot x} \,
\bigl\langle 0 \bigl| \mathrm{Tr}\!\bigl[ \gamma_5 S_s^{ee'}(x) \gamma_5 C{S}_q^{dd' \mathrm{T}}(x)C \bigr] \nonumber \\
& \times \mathrm{Tr}\!\bigl[ \gamma_5 S_c^{gg'}(x) \gamma_5 C{S}_s^{ff' \mathrm{T}}C(x) \bigr] 
\, C{S}_c^{c'c \mathrm{T}}C(-x) \bigr| 0 \bigr\rangle_{\gamma},\\
\Pi^{\mathrm{QCD}}_{\mu\nu, 1}(p,q) &= -i \, \varepsilon^{abc}\varepsilon^{a'b'c'}\varepsilon^{ade}\varepsilon^{a'd'e'}\varepsilon^{bfg} \varepsilon^{b'f'g'} 
\int d^4x \, e^{ip\cdot x} \, \bigl\langle 0 \bigl|
\mathrm{Tr}\!\bigl[ \gamma_\mu S_c^{gg'}(x) \gamma_\nu C{S}_s^{ff' \mathrm{T}}C(x) \bigr] 
\nonumber \\
& \times
 \mathrm{Tr}\!\bigl[ \gamma_5 S_s^{ee'}(x) \gamma_5 C{S}_q^{dd' \mathrm{T}}(x)C \bigr]  
\, C{S}_c^{c'c \mathrm{T}}C(-x) \bigr| 0 \bigr\rangle_{\gamma}.\label{eq:qcd_explicit1}
\end{align}
In these expressions, the subscript \(\gamma\) on the vacuum expectation value signifies that the electromagnetic field is included through its interactions with the quark propagators. 
The light (\(q = u,d,s\)) and charm quark propagators in coordinate space are fundamental building blocks. To leading order in the strong coupling \(g_s\), the light-quark propagator can be expressed as \cite{Balitsky:1987bk}:
\begin{align}
\label{eq:light_propagator}
S_q^{ab}(x) &= \frac{i \delta^{ab}}{2\pi^2 x^4} \slashed{x} 
- \frac{m_q \delta^{ab}}{4\pi^2 x^2}
- \frac{i g_s}{16\pi^2 x^2} \int_0^1 dv \, G_{\mu\nu}^{ab}(vx) 
\bigl[ \bar{v} \slashed{x} \sigma^{\mu\nu} + v \sigma^{\mu\nu} \slashed{x} \bigr],
\end{align}
where  \(G_{\mu\nu}^{ab}\) is the gluon field strength tensor in the adjoint representation. 
For the charm quark, the propagator assumes a different asymptotic form \cite{Belyaev:1985wza}:
\begin{align}
\label{eq:charm_propagator}
S_c^{ab}(x) &= \frac{m_c^2 \delta^{ab}}{4\pi^2} \Biggl[ 
\frac{K_1\bigl(m_c\sqrt{-x^2}\bigr)}{\sqrt{-x^2}} 
+ i \frac{\slashed{x} \, K_2\bigl(m_c\sqrt{-x^2}\bigr)}{(\sqrt{-x^2})^2} \Biggr] 
- i \frac{g_s m_c}{16\pi^2} \int_0^1 dv \, G_{\mu\nu}^{ab}(vx) 
\Biggl[ \frac{K_1\bigl(m_c\sqrt{-x^2}\bigr)}{\sqrt{-x^2}} 
\bigl( \sigma^{\mu\nu} \slashed{x} + \slashed{x} \sigma^{\mu\nu} \bigr)\nonumber \\
& 
+ 2 \sigma^{\mu\nu} K_0\bigl(m_c\sqrt{-x^2}\bigr) \Biggr],
\end{align}
where \(K_n(z)\) are modified Bessel functions of the second kind.  
We note that the quark condensates and the gluon condensate do not appear explicitly in the propagator expressions above. They enter the calculation through the non-perturbative soft photon coupling
described in Eq.~(\ref{eq:nonperturbative_insertion}) below, which introduces these condensates via the photon DAs.  

The interaction of the external photon with the quark lines is incorporated via two distinct mechanisms, corresponding to different regimes of the photon momentum fraction. 
Perturbative (hard) photon emission: Here, the photon couples directly to a quark line through an electromagnetic vertex. Mathematically, this is implemented by replacing one of the free propagators in the correlation function with its perturbatively expanded form that includes one photon insertion:
\begin{equation}
\label{eq:perturbative_insertion}
S^{\mathrm{free}}(x) \;\longrightarrow\; \int d^4z \, S^{\mathrm{free}}(x-z) \, \slashed{A}(z) \, S^{\mathrm{free}}(z),
\end{equation}
where \(A_\mu(z)\) represents the photon field and \(S^{\mathrm{free}}(x)\) denotes the leading (free) part of either the light or heavy quark propagator. Non-perturbative (soft) photon coupling: When the photon carries a small momentum fraction, its interaction with the quark fields is more appropriately described through photon DAs. This is achieved by substituting one of the light-quark propagators with a non-local matrix element that encodes the quark-antiquark component of the photon wave function:
\begin{equation}
\label{eq:nonperturbative_insertion}
S_{\mu\nu}^{ab}(x) \;\longrightarrow\; -\frac{1}{4} \bigl[ \bar{q}^a(x) \Gamma_i q^b(0) \bigr] \bigl( \Gamma_i \bigr)_{\mu\nu},
\end{equation}
with \(\Gamma_i\) running over the complete basis of Dirac matrices: \(1, \gamma_5, \gamma_\alpha, i\gamma_5\gamma_\alpha, \sigma_{\alpha\beta}/2\). 

The matrix elements \(\langle \gamma(q) | \bar{q}(x) \Gamma_i q(0) | 0 \rangle\) and \(\langle \gamma(q) | \bar{q}(x) \Gamma_i G_{\mu\nu} q(0) | 0 \rangle\) are then expressed in terms of photon DAs following the established formalism \cite{Ball:2002ps}. After inserting the quark propagators Eqs.~(\ref{eq:light_propagator}) and (\ref{eq:charm_propagator}) into Eqs.~(\ref{eq:qcd_explicit})-(\ref{eq:qcd_explicit1}) and implementing the photon couplings through both mechanisms Eqs.~(\ref{eq:perturbative_insertion}) and (\ref{eq:nonperturbative_insertion}), we arrive at the coordinate-space representation of the correlation function as a sum of diverse contributions: perturbative photon emission from quark lines, non-perturbative contributions from photon DAs, and mixed terms. The final step consists of performing the Fourier transformation to momentum space:
\begin{equation}
\label{eq:fourier_transform}
\Pi^{\mathrm{QCD}}(p,q) = \int d^4x \, e^{ip\cdot x} \, \mathcal{F}(x; q),
\end{equation}
where \(\mathcal{F}(x; q)\) symbolizes the coordinate-space expression obtained from the OPE. The resulting momentum-space correlation function comprises terms proportional to various Lorentz structures, whose coefficients are given by integrals over the photon DAs and the momentum fraction variables. These coefficients will be matched to the corresponding structures in the phenomenological representation to derive the sum rules for the magnetic dipole moments. The analogous procedure is applied to all other interpolating currents \(J_2(x)\), \(J_3(x)\), \(J_4(x)\) for spin-1/2 and \(J_\mu^1(x)\), \(J_\mu^2(x)\), \(J_\mu^3(x)\), \(J_\mu^4(x)\) for spin-3/2, yielding similar but technically distinct expressions that mirror the different Dirac structures inherent in these currents.

\subsection{Derivation of the sum rules for the magnetic dipole moments}
\label{subsec:sum_rules}

With the hadronic and QCD representations of the correlation functions established, we now proceed to derive the sum rules for the magnetic dipole moments. The fundamental principle of QCD sum rules is to equate the two representations through the quark-hadron duality and then apply a Borel transformation to suppress contributions from higher states and enhance the convergence of the operator product expansion. After isolating the coefficients of the relevant Lorentz structures on both the hadronic and QCD sides, and applying the standard Borel transformation procedure with respect to the variable \(-p^2\) and \(-(p+q)^2\), we obtain the following master sum rules for the magnetic dipole moments of the doubly strange hidden-charm pentaquarks:
\begin{align}
\label{eq:sum_rule_12}
\mu_{P^{\Lambda}_{\psi ss}}(J_i) \, \lambda_{P^{\Lambda}_{\psi ss}}^2(J_i) \, 
e^{-m^2(J_i)/M^2} &= \Pi_i(M^2, s_0), \\[6pt]
\label{eq:sum_rule_32}
\mu_{P^{\ast\Lambda}_{\psi ss}}(J_\mu^i) \, \lambda_{P^{\ast\Lambda}_{\psi ss}}^2(J_\mu^i) \, 
e^{-{m^\ast}^2(J_\mu^i)/M^2} &= \Xi_i(M^2, s_0),
\end{align}
where \(i = 1, 2, 3, 4\) labels the four distinct interpolating currents for each spin sector. In these expressions, \(M^2\) denotes the Borel mass parameter, \(s_0\) is the continuum threshold that separates the ground-state contribution from higher resonances and the continuum. The functions \(\Pi_i(M^2, s_0)\) and \(\Xi_i(M^2, s_0)\) encapsulate the results of the QCD side calculation after Borel transformation and continuum subtraction, and they depend on various input parameters including quark masses, quark condensates, and  photon DAs.

The explicit forms of  \(\Pi_i(M^2, s_0)\) and \(\Xi_i(M^2, s_0)\) are obtained by carrying out the lengthy operator product expansion described in the previous subsection. Given the considerable complexity and structural similarity of the analytical expressions for all eight currents, we present here the complete result only for the first spin-1/2 current \(J_1(x)\) as a representative example. The corresponding expressions for the other currents (\(J_2(x), J_3(x), J_4(x)\) and \(J_\mu^1(x), J_\mu^2(x), J_\mu^3(x), J_\mu^4(x)\)) are derived following an identical computational procedure and exhibit analogous analytical structures, differing primarily in the numerical coefficients arising from the different Dirac matrix arrangements. The magnetic dipole moment for a given current is then extracted by solving the sum rule: 
\begin{align}
\label{eq:mu_extract_12}
\mu_{P^{\Lambda}_{\psi ss}}(J_i) &= 
\frac{\Pi_i(M^2, s_0)}{\lambda_{P^{\Lambda}_{\psi ss}}^2(J_i)} \, 
e^{m^2(J_i)/M^2}, \\
\mu_{P^{\ast\Lambda}_{\psi ss}}(J_\mu^i) &= 
\frac{\Xi_i(M^2, s_0)}{\lambda_{P^{\ast\Lambda}_{\psi ss}}^2(J_\mu^i)} \, 
e^{{m^\ast}^2(J_\mu^i)/M^2}.
\end{align}

The pole residues \(\lambda(J)\) and masses \(m(J)\) that appear in these formulas are not free parameters; they are determined from the mass sum rules for the same interpolating currents within the same QCD sum rule framework \cite{Wang:2025pjt}.  To ensure consistency, we use the values obtained from their analysis of the spectroscopic properties of these states, where the working regions for \(M^2\) and \(s_0\) were established based on standard sum rule stability criteria.

In the following section, we will specify the numerical values of all input parameters and discuss the determination of the optimal working windows for \(M^2\) and \(s_0\). The explicit analytical expression for \(\Pi_1(M^2, s_0)\) corresponding to the current \(J_1(x)\) is provided in the Appendix.

\end{widetext}

\section{Numerical Analysis} \label{sec:numerical}

\subsection{Numerical inputs and stability of LCSR}

This section presents the numerical analysis of the LCSR for the electromagnetic multipole moments of the doubly strange hidden-charm pentaquarks. For the numerical evaluation, we specify the input QCD parameters, determine the working regions of the Borel mass and continuum threshold, and then extract the physical observables together with their uncertainties. The input parameters are taken from well-established sources in the literature. We use the following values: 
\begin{itemize}
    \item Strange quark mass: 
    $m_s = 93.5 \pm 0.8~\text{MeV}$~\cite{ParticleDataGroup:2024cfk}
    \item Charm quark mass: 
    $m_c = 1.273 \pm 0.0046~\text{GeV}$~\cite{ParticleDataGroup:2024cfk}
    \item Light quark condensate: 
    $\langle \bar qq \rangle = (-0.24 \pm 0.01)^3~\text{GeV}^3$~\cite{Ioffe:2005ym}
    \item Strange quark condensate: 
    $\langle \bar ss \rangle = (0.8 \pm 0.1) \times \langle \bar qq \rangle$~\cite{Ioffe:2005ym}
    \item Gluon condensate: 
    $\langle g_s^2 G^2 \rangle = 0.48 \pm 0.14~\text{GeV}^4$~\cite{Narison:2018nbv}
    \item Nonperturbative constant: 
    $f_{3\gamma} = -0.0039~\text{GeV}^2$~\cite{Ball:2002ps}
    \item Magnetic susceptibility: 
    $\chi = -2.85 \pm 0.5~\text{GeV}^{-2}$~\cite{Rohrwild:2007yt}
\end{itemize}
For the light quarks we set $m_u = m_d = 0$ and retain only terms linear in $m_s$, while contributions of order $m_s^2$ are neglected due to their negligible numerical impact. Photon DAs are adopted from~\cite{Ball:2002ps}, and contributions up to twist-4 are included. The masses and pole residues are taken from~\cite{Wang:2025pjt}, where they were determined using mass sum rules with the same interpolating currents.

The sum rule predictions depend on two auxiliary parameters: the Borel mass $M^2$ and the continuum threshold $s_0$. In principle, physical observables should be independent of these parameters, but in practice some residual dependence remains. We therefore determine optimal working regions where the results exhibit minimal sensitivity to $M^2$ and $s_0$, guided by two standard criteria:

\paragraph{Pole dominance (PC):} The contribution of the ground state should dominate over that of higher states and the continuum. We require
\begin{equation}
\label{eq:pole_criterion}
\text{PC} = \frac{\Pi_i(M^2, s_0)}{\Pi_i(M^2, \infty)} \geq 40\%,
\end{equation}
where $\Pi_i(M^2, \infty)$ represents the sum rule without continuum subtraction.

\paragraph{OPE convergence (CVG):} The higher-dimensional terms in the OPE should be sufficiently suppressed. We demand that the highest-dimensional contributions (dimension-7 in our analysis, corresponding to mixed condensate-gluon condensate terms) satisfy
\begin{equation}
\label{eq:convergence_criterion}
\text{CVG} = \frac{\Pi_i^{\text{Dim7}}(M^2, s_0)}{\Pi_i(M^2, s_0)} < 5\%.
\end{equation}

\begin{figure}[htbp]
\centering
\includegraphics[width=0.45\textwidth]{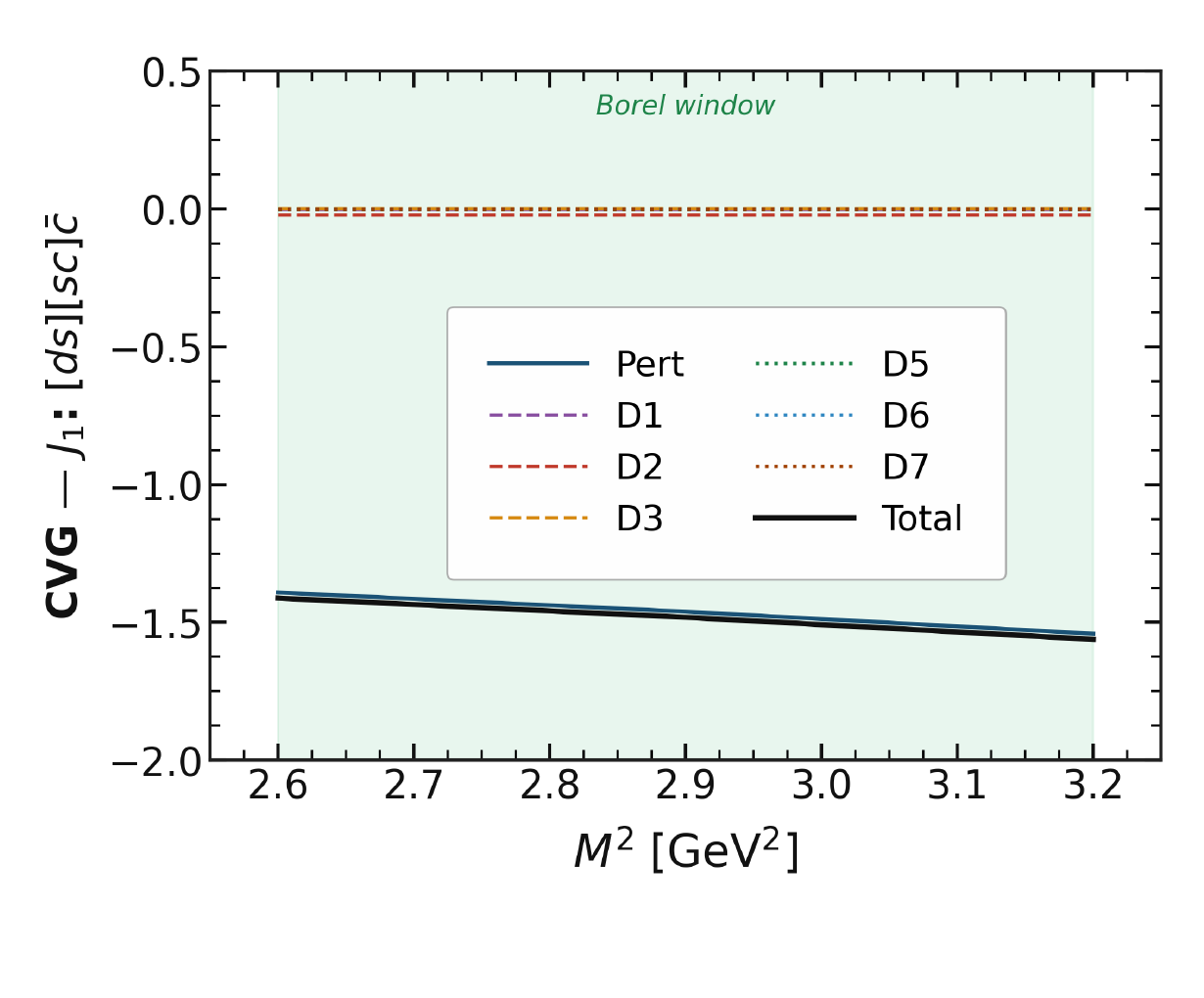} \qquad
\includegraphics[width=0.45\textwidth]{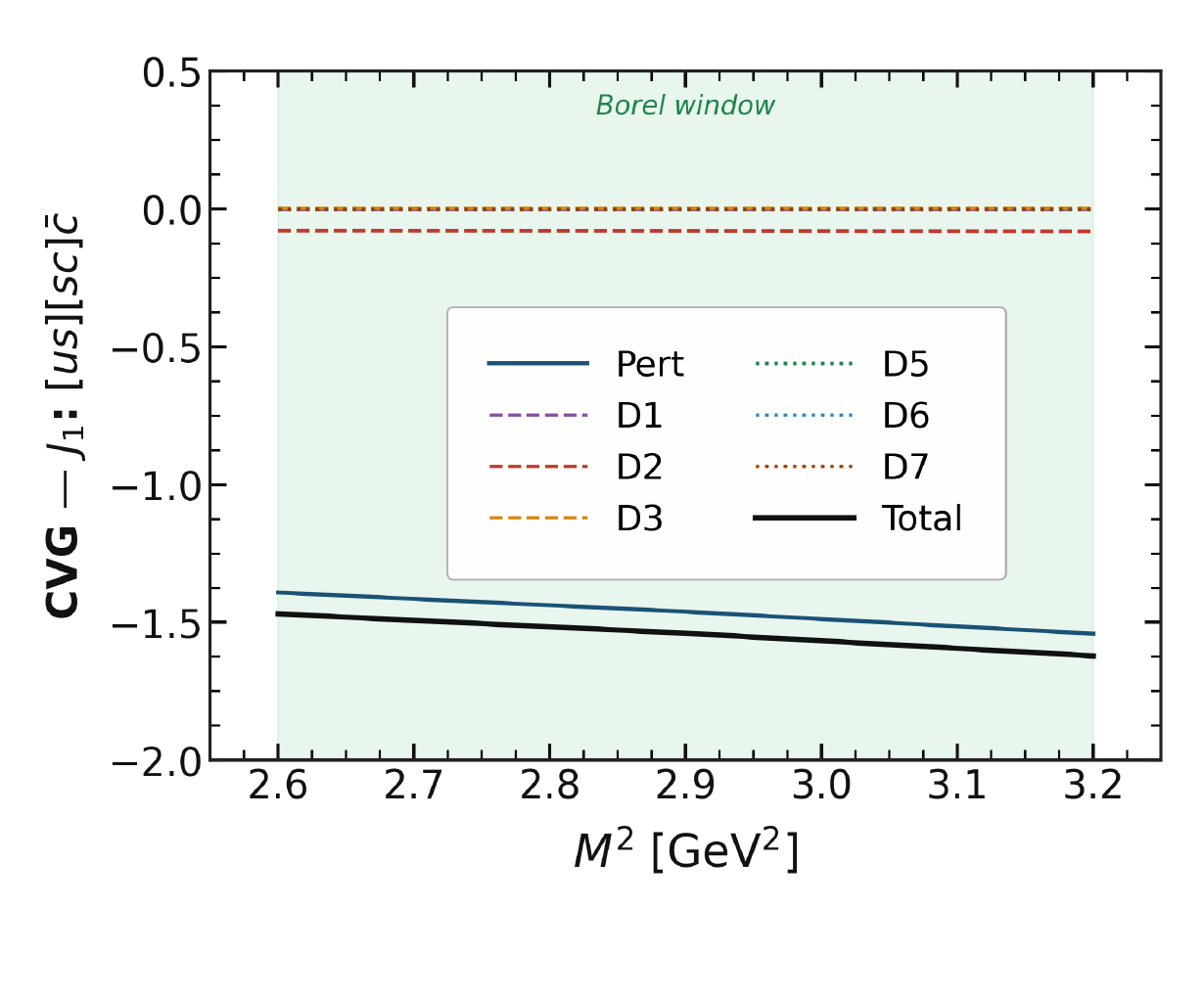}\\
\includegraphics[width=0.45\textwidth]{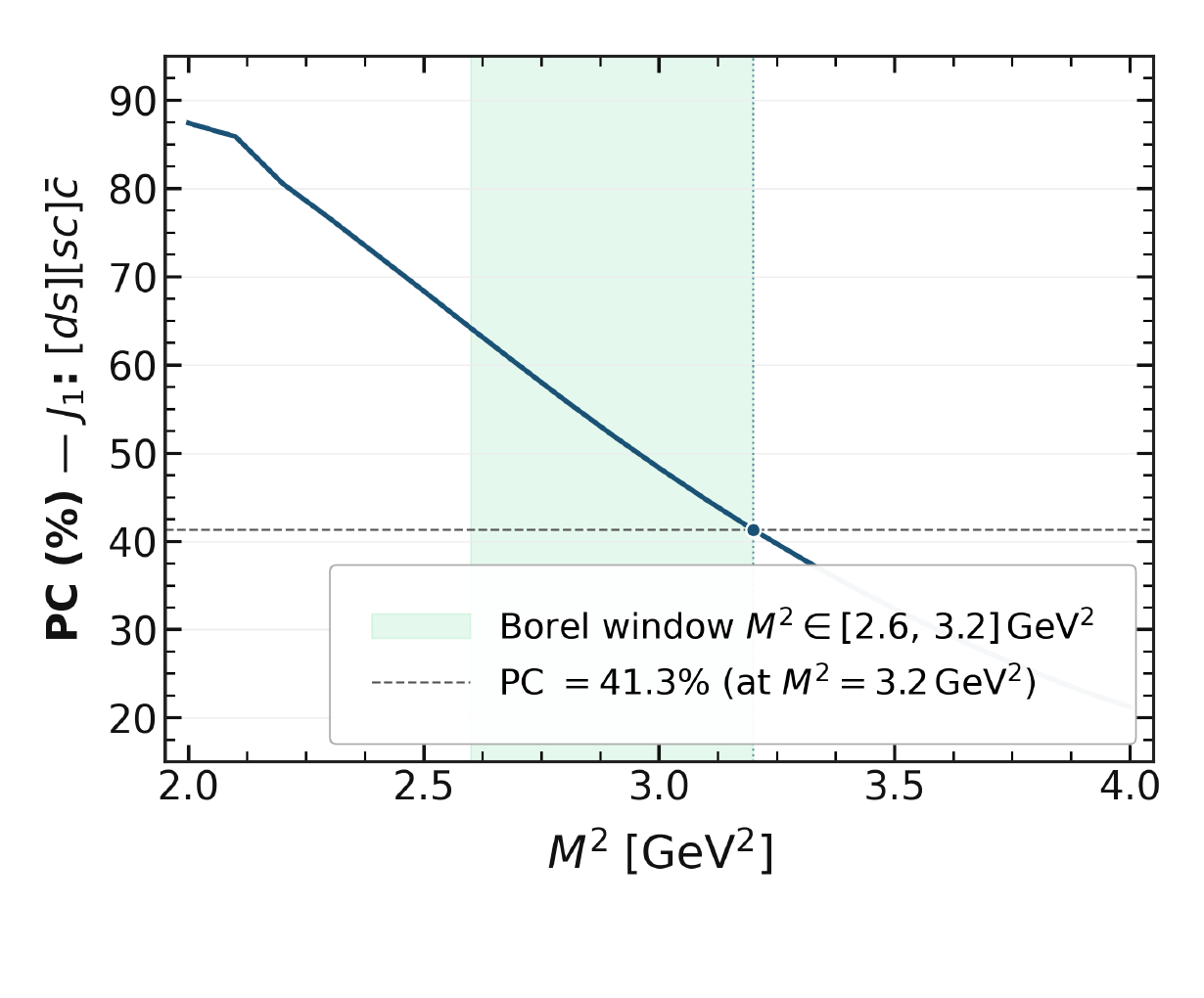} \qquad
\includegraphics[width=0.45\textwidth]{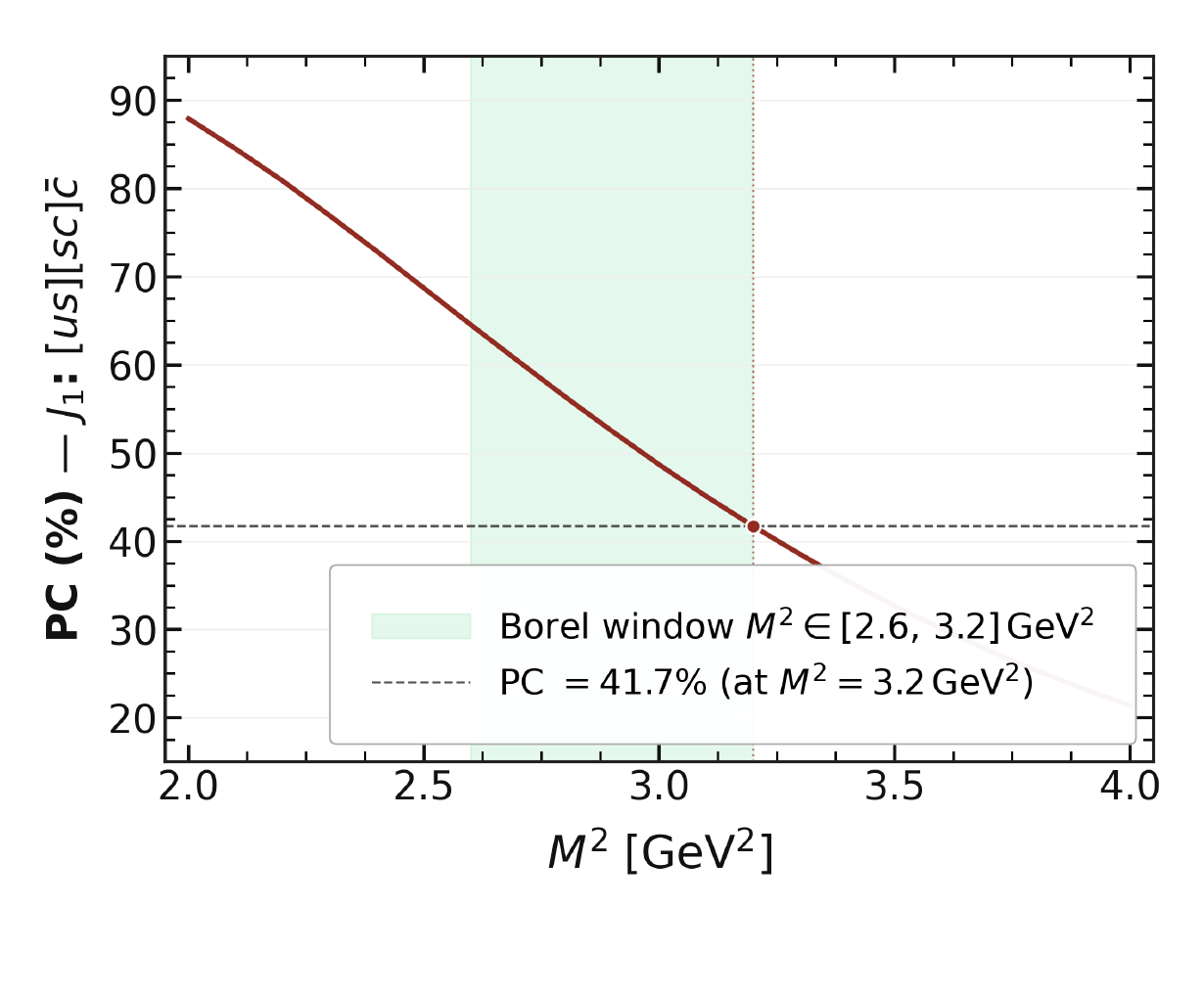}\\
\includegraphics[width=0.45\textwidth]{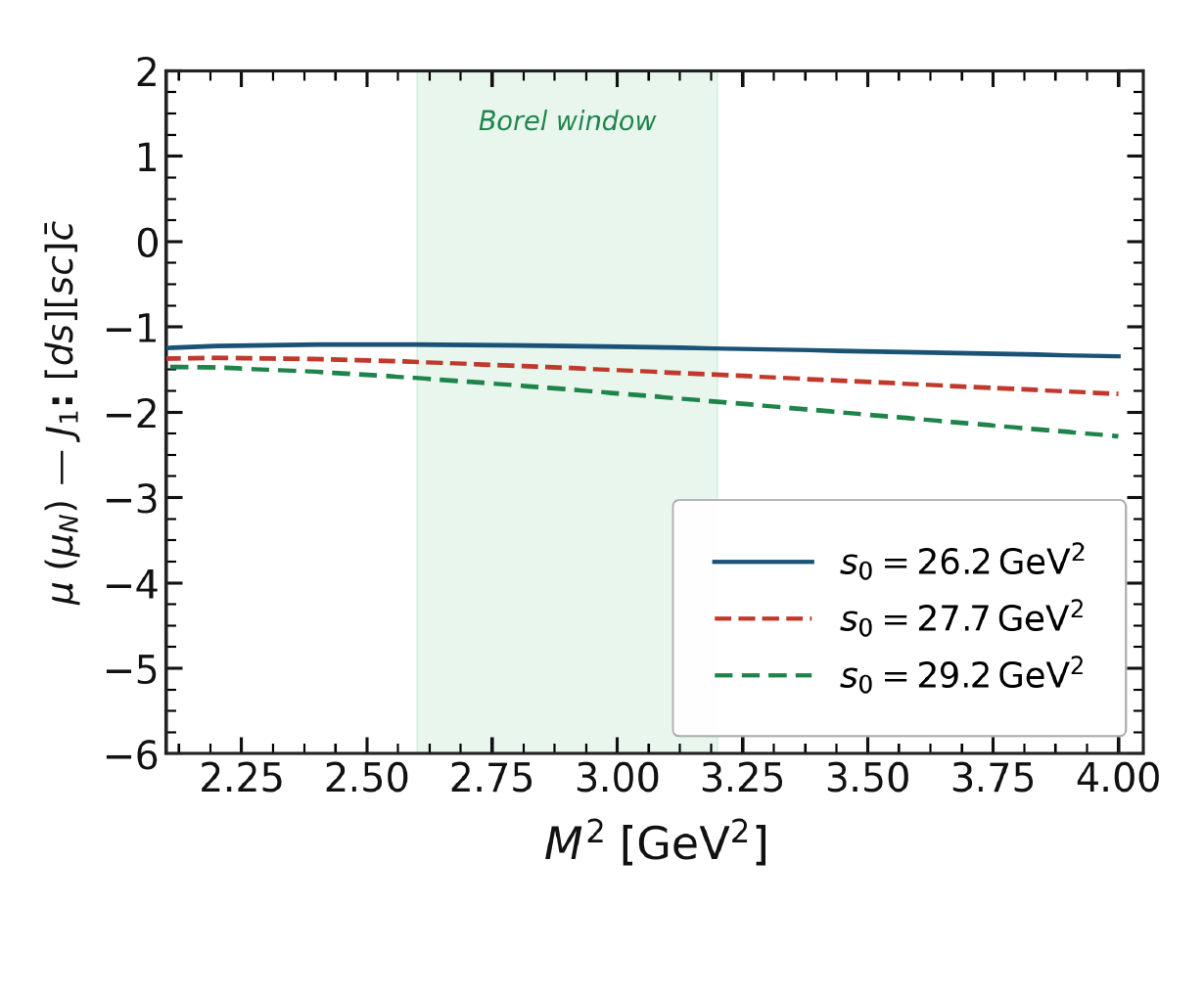} \qquad
\includegraphics[width=0.45\textwidth]{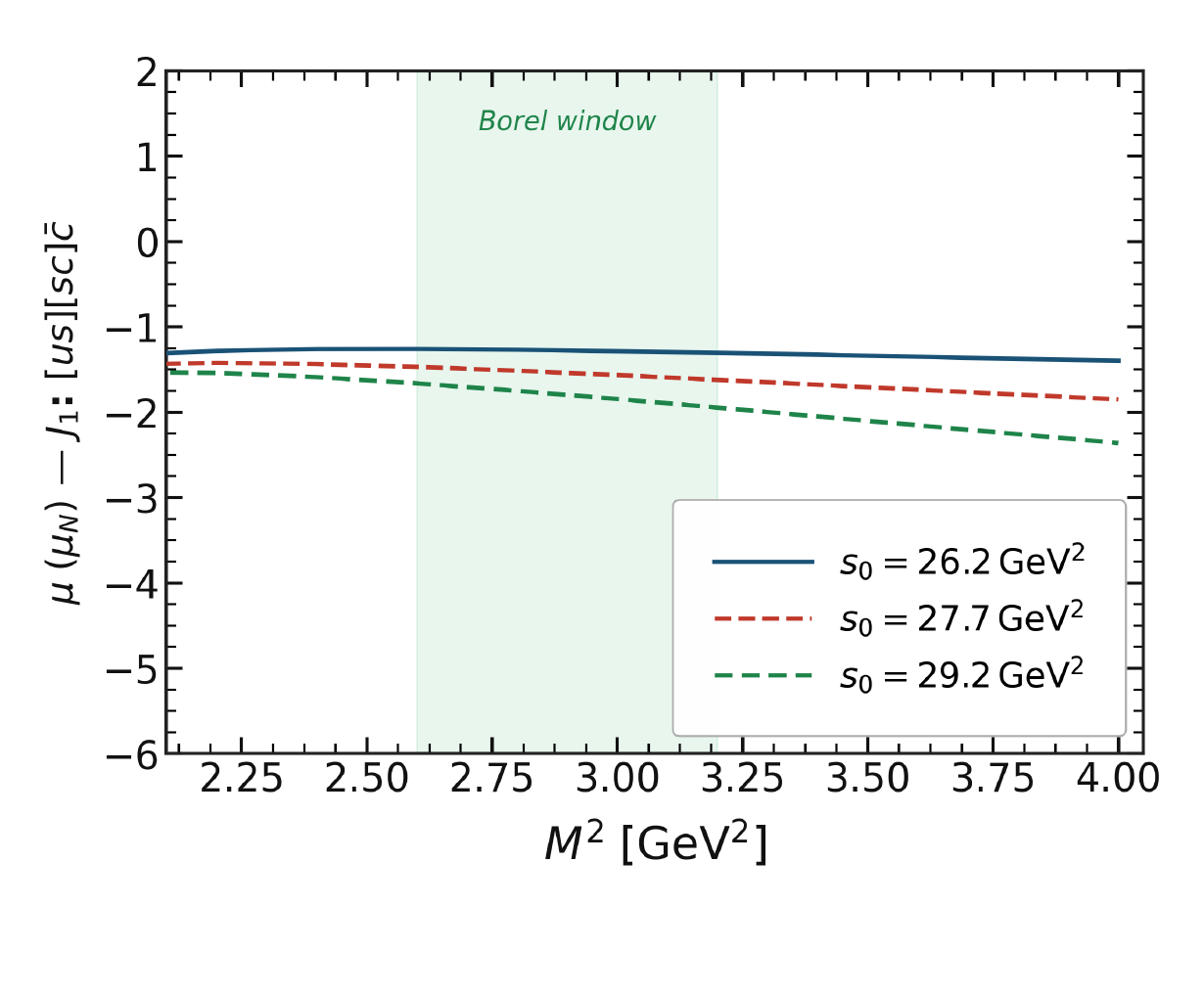}\\
\caption{Sum rule analysis for the $J_1$ current in the $[ds][sc]\bar{c}$ (left column) and $[us][sc]\bar{c}$ (right column) configurations. Top row: Convergence of the operator product expansion (OPE) showing 
the contributions of individual condensate terms as a function of the Borel mass $M^2$. Middle row: Pole contribution (PC) as a function of $M^2$; the horizontal dashed line marks the PC value at the upper 
boundary of the Borel window ($M^2 = 3.2\,\mathrm{GeV}^2$), and the shaded band indicates the selected working region $M^2 \in [2.6,\,3.2]\,\mathrm{GeV}^2$. Bottom row: Magnetic dipole moment $\mu\,(\mu_N)$ as a function of $M^2$ for three values of the continuum threshold $s_0$, demonstrating stability within the Borel window.}
\label{fig:analysis_J1}
\end{figure}

Our OPE includes contributions from various operator dimensions: dimension-1 ($\langle \bar q q \rangle \chi$, $\langle \bar s s \rangle \chi$), dimension-2 ($f_{3\gamma}$), dimension-3 ($\langle \bar q q \rangle$, $\langle \bar s s \rangle$), dimension-5 ($\langle g_s^2 G^2 \rangle \langle \bar q q \rangle \chi$, $\langle g_s^2 G^2 \rangle \langle \bar s s \rangle \chi$), dimension-6 ($\langle g_s^2 G^2 \rangle f_{3\gamma}$), and dimension-7 ($\langle g_s^2 G^2 \rangle \langle \bar q q \rangle$, $\langle g_s^2 G^2 \rangle \langle \bar s s \rangle$). Applying these criteria separately for each interpolating current, we determine the corresponding working regions of $M^2$ and $s_0$, summarized in Table~\ref{tab:working_regions}. 
A further comment is in order regarding the parity content of the local interpolating currents employed here. Such currents do not, in general, project onto a single parity eigenstate and can in principle couple to both negative- and positive-parity pentaquarks sharing the same spin and flavor quantum numbers. The positive-parity partner is expected to lie $300$--$500$~MeV above the negative-parity ground state due to the additional $P$-wave excitation required by parity flipping, a hierarchy consistent with QCD sum rule analyses of related hidden-charm and doubly-charm pentaquark systems~\cite{Pimikov:2019dyr, Duan:2024uuf}. In the present sum-rule framework, the contribution of this partner is suppressed by the Borel exponential factor $e^{-m^2/M^2}$, which disfavors higher-mass states: within our Borel window, the ratio $e^{-m_+^2/M^2}/e^{-m_-^2/M^2}$ evaluates to $\approx 0.2$--$0.3$, corresponding to a suppression by a factor of roughly $4$--$5$. The stability of the extracted moments within the working windows of $M^2$ and $s_0$ (Table~\ref{tab:working_regions} and Fig.~\ref{fig:analysis_J1}), together with pole contributions
reaching $60$--$70\%$ at the lower end of the Borel range, provides further empirical evidence that the dominant contribution in this region is a single ground-state pole. The residual parity contamination is therefore quantitatively subdominant to the other uncertainties discussed below. 
The CVG and PC are evaluated numerically for every state. The dimension-7 contributions remain well below the imposed 5\% bound and are explicitly listed in Table~\ref{tab:working_regions}, where they are seen to stay at the sub-percent level for all currents. Meanwhile, the pole contribution lies between 40\% and 70\% within the selected Borel windows, ensuring a reliable dominance of the ground-state signal. 
Figure~\ref{fig:analysis_J1} illustrates these criteria for the current $J_1(x)$ as a representative example. The left panel shows the CVG as a function of $M^2$, confirming that the dimension-7 contributions remain negligible throughout the Borel window. The middle panel displays the PC, with the horizontal line indicating the minimum value  and vertical lines marking the selected Borel region where this condition is satisfied. The right panel presents the magnetic dipole moment itself, demonstrating its stability within the chosen working region. 

\begin{table}[htbp]
\centering
\caption{Working regions for the continuum threshold $s_0$ and Borel
parameter $M^2$, together with the OPE convergence (CVG) and pole
contribution (PC) for each interpolating current.}
\label{tab:working_regions}
\renewcommand{\arraystretch}{1.25}
\setlength{\tabcolsep}{10pt}
\begin{tabular}{lcccc}
\toprule
Current & $s_0\;(\mathrm{GeV}^2)$ & $M^2\;(\mathrm{GeV}^2)$
        & CVG (\%) & PC (\%) \\
\midrule
\multicolumn{5}{l}{\textit{Spin-$\frac{1}{2}$ sector}} \\[3pt]
$J_1(x)$       & $[26.2,\;29.2]$ & $[2.6,\;3.2]$ & $0.61$ & $[64.26,\;41.31]$ \\
$J_2(x)$       & $[26.2,\;29.2]$ & $[2.8,\;3.4]$ & $0.68$ & $[62.55,\;41.18]$ \\
$J_3(x)$       & $[26.6,\;29.2]$ & $[2.7,\;3.3]$ & $0.86$ & $[64.43,\;42.22]$ \\
$J_4(x)$       & $[26.2,\;29.2]$ & $[2.6,\;3.2]$ & $0.54$ & $[64.82,\;41.80]$ \\
\midrule
\multicolumn{5}{l}{\textit{Spin-$\frac{3}{2}$ sector}} \\[3pt]
$J_\mu^2(x)$   & $[26.2,\;29.2]$ & $[2.6,\;3.2]$ & $0.59$ & $[62.85,\;40.58]$ \\
$J_\mu^3(x)$   & $[26.6,\;29.8]$ & $[2.7,\;3.5]$ & $0.72$ & $[67.95,\;40.22]$ \\
$J_\mu^4(x)$   & $[26.6,\;29.8]$ & $[2.7,\;3.5]$ & $0.51$ & $[70.39,\;42.60]$ \\
\bottomrule
\end{tabular}
\end{table}

With all parameters fixed, we now present our predictions for the electromagnetic multipole moments of the doubly strange hidden-charm pentaquarks. Table~\ref{tab:multipole_results} summarizes the magnetic dipole moments (in nuclear magnetons $\mu_N$) for both $[ds][sc]\bar c$ and $[us][sc]\bar c$ configurations across all interpolating currents. For the spin-$3/2$ states, we additionally provide the electric quadrupole ($\mathcal{Q}$ in $10^{-2}$~fm$^2$) and magnetic octupole ($\mathcal{O}$ in $10^{-3}$~fm$^3$) moments where applicable. For the current $J^1_\mu(x)$, no sum rule for the spin-$\frac{3}{2}$ magnetic dipole moment can be constructed. When the OPE is evaluated for this current, the QCD side of the correlation function does not produce the Lorentz structures identified in Eq.~(\ref{eq:phen_explicit_32}) that are necessary to
match the hadronic representation and isolate the spin-$\frac{3}{2}$ form factors $F_1$ and $F_2$. This is a structural consequence of the scalar--scalar $[C\gamma_5][C\gamma_5]$ diquark configuration of $J^1_\mu(x)$:
the resulting operator couples dominantly to the spin-$\frac{1}{2}$ sector, leaving the spin-$\frac{3}{2}$ projection undetermined. Unlike the other excluded cases which may suffer from numerical instability, this exclusion is categorical---no choice of Borel parameter or continuum threshold can resolve it.
We therefore exclude $J^1_\mu(x)$ from the spin-$\frac{3}{2}$ analysis entirely. The quoted uncertainties include the errors of the input parameters (quark masses, condensates, and other nonperturbative inputs), the variation of the $M^2$ and $s_0$ within their working regions, as well as the uncertainties of the pole residues and masses taken from our previous spectroscopic study. The uncertainty budget of our predictions is dominated by three sources. Variations of the continuum threshold $s_0$ within its working region account for approximately $27\%$ of the total error, reflecting the residual sensitivity of the sum rules to the modeling of higher-state contributions. The pole residues and pentaquark masses imported from~\cite{Wang:2025pjt} contribute $22\%$ and $18\%$, respectively; these quantities are varied independently within their quoted $1\sigma$ ranges, as the correlations between them within the original spectroscopic analysis are not available to us. The remaining uncertainty is distributed among the photon DAs ($\sim 10\%$), the light and charm quark masses ($\sim 10\%$), the Borel parameter $M^2$ ($\sim 8\%$), and the QCD condensate inputs ($\sim 5\%$). Each source is treated as independent of the others, and all contributions are added in quadrature to obtain the total uncertainties reported in Table~\ref{tab:multipole_results}.

\begin{table}[htbp]
\centering
\caption{Predictions for the electromagnetic multipole moments of
$P^{\Lambda}_{\psi ss}$ pentaquarks. Magnetic dipole moments are given
in nuclear magnetons $\mu_N$, electric quadrupole moments in
$10^{-2}$~fm$^2$, and magnetic octupole moments in $10^{-3}$~fm$^3$.
Dashes indicate moments not defined for spin-$\frac{1}{2}$ states.}
\label{tab:multipole_results}
\renewcommand{\arraystretch}{1.25}
\setlength{\tabcolsep}{9pt}
\begin{tabular}{llccc}
\toprule
& & Magnetic Dipole
  & Electric Quadrupole
  & Magnetic Octupole \\
Current & Configuration
  & $\mu_{P^{\Lambda}_{\psi ss}}$
  & $\mathcal{Q}_{P^{\Lambda}_{\psi ss}}$
  & $\mathcal{O}_{P^{\Lambda}_{\psi ss}}$ \\
\midrule
\multicolumn{5}{l}{\textit{Spin-$\tfrac{1}{2}$ states}} \\[3pt]
\multirow{2}{*}{$J_1(x)$}
  & $[ds][sc]\bar{c}$ & $-1.48 \pm 0.38$ & \multirow{2}{*}{---} & \multirow{2}{*}{---} \\
  & $[us][sc]\bar{c}$ & $-1.54 \pm 0.39$ & & \\[4pt]
\multirow{2}{*}{$J_2(x)$}
  & $[ds][sc]\bar{c}$ & $\phantom{-}4.12 \pm 1.03$ & \multirow{2}{*}{---} & \multirow{2}{*}{---} \\
  & $[us][sc]\bar{c}$ & $\phantom{-}4.16 \pm 1.04$ & & \\[4pt]
\multirow{2}{*}{$J_3(x)$}
  & $[ds][sc]\bar{c}$ & $\phantom{-}5.62 \pm 1.41$ & \multirow{2}{*}{---} & \multirow{2}{*}{---} \\
  & $[us][sc]\bar{c}$ & $\phantom{-}5.74 \pm 1.44$ & & \\[4pt]
\multirow{2}{*}{$J_4(x)$}
  & $[ds][sc]\bar{c}$ & $-2.06 \pm 0.52$ & \multirow{2}{*}{---} & \multirow{2}{*}{---} \\
  & $[us][sc]\bar{c}$ & $-2.15 \pm 0.54$ & & \\
\midrule
\multicolumn{5}{l}{\textit{Spin-$\tfrac{3}{2}$ states}} \\[3pt]
\multirow{2}{*}{$J_\mu^2(x)$}
  & $[ds][sc]\bar{c}$ & $-0.43 \pm 0.11$ & $-1.40 \pm 0.35$ & $-0.26 \pm 0.07$ \\
  & $[us][sc]\bar{c}$ & $-0.48 \pm 0.12$ & $-1.40 \pm 0.35$ & $-0.26 \pm 0.07$ \\[4pt]
\multirow{2}{*}{$J_\mu^3(x)$}
  & $[ds][sc]\bar{c}$ & $-1.14 \pm 0.29$ & $\phantom{-}3.52 \pm 0.88$ & $-0.96 \pm 0.24$ \\
  & $[us][sc]\bar{c}$ & $-3.66 \pm 0.91$ & $-2.01 \pm 0.50$ & $\phantom{-}0.00 \pm 0.00$ \\[4pt]
\multirow{2}{*}{$J_\mu^4(x)$}
  & $[ds][sc]\bar{c}$ & $-4.25 \pm 1.06$ & $\phantom{-}5.55 \pm 1.39$ & $-1.51 \pm 0.38$ \\
  & $[us][sc]\bar{c}$ & $-4.13 \pm 1.03$ & $\phantom{-}5.55 \pm 1.39$ & $-1.51 \pm 0.38$ \\
\bottomrule
\end{tabular}
\end{table}


\subsection{Discussion of the electromagnetic moments}\label{subsec:discussion}

The electromagnetic moments presented in Table~\ref{tab:multipole_results} exhibit a striking and systematic dependence on the choice of interpolating current, offering a direct window into the internal dynamics of these exotic states. The magnetic dipole moments span an exceptionally broad range, from $-4.25\,\mu_N$ to $+5.74\,\mu_N$---a spread of nearly $10\,\mu_N$ that underscores the extreme sensitivity of this observable to the underlying quark-gluon structure.

For the spin-$1/2$ sector, a clear dichotomy emerges based on the diquark configuration. Currents built from axial-vector diquarks ($J_2(x)$, $J_3(x)$) yield positive moments of order $4$--$6\,\mu_N$, whereas those with scalar diquarks ($J_1(x)$, $J_4(x)$) produce negative values around $-2\,\mu_N$. This sharp contrast provides a powerful diagnostic: a future experimental measurement of the magnetic moment could, by its very sign, discriminate between competing internal configurations. A positive moment would point decisively toward axial-vector diquark correlations, while a negative value would favor a scalar diquark picture.

The spin-$3/2$ moments are predominantly negative across all currents, with $J_\mu^4(x)$ yielding the largest magnitude at $-4.25\,\mu_N$. While isospin symmetry is generally respected---differences between $[ds][sc]\bar c$ and $[us][sc]\bar c$ configurations typically remain below $5\%$---the $J_\mu^3(x)$ current stands out as a notable exception. Here, the light-quark substitution induces a dramatic shift from $-1.14\,\mu_N$ to $-3.66\,\mu_N$, revealing a pronounced sensitivity whose physical origin is analyzed in detail below.

The origin of this systematic behavior becomes transparent upon decomposing the moments into individual quark-flavor contributions, as shown in Table~\ref{tab:quark_magnetic}. The charm quark dominates in almost every configuration, its large electric charge and reduced relativistic suppression making it the primary engine of the magnetic response. The strange quark, by contrast, plays a remarkably variable role: it contributes sizably and positively in axial-vector diquark currents, but remains suppressed---often negative---in scalar ones. This duality directly reflects the spin correlations encoded in the diquark structures: axial-vector diquarks (spin 1) mobilize the strange quark's magnetic potential, while scalar diquarks (spin 0) effectively neutralize it. Light quarks contribute only marginally and respect isospin symmetry in all currents except $J^3_\mu(x)$. In this current, the light quark occupies an axial-vector diquark position $[qs]$ coupled via $C\gamma_\mu$, which isolates it from the spin averaging that suppresses its contribution in scalar diquark structures. As a result, the $d\to u$ substitution acts through two simultaneous effects: the charge ratio $e_u/e_d = -2$ and a reversal of the spin orientation, yielding $\mu_u/\mu_d = -2.00$---a value confirmed exactly by Table~\ref{tab:quark_magnetic}. This behavior is not an anomaly but a direct and transparent consequence of the Dirac structure of $J^3_\mu(x)$.


\begin{table}[htbp]
\centering
\caption{Quark-level contributions to the magnetic dipole moments
(in units of $\mu_N$) for the $P_{\psi ss}^{\Lambda}$ pentaquarks
in both $[ds][sc]\bar{c}$ and $[us][sc]\bar{c}$ configurations.}
\label{tab:quark_magnetic}
\renewcommand{\arraystretch}{1.25}
\setlength{\tabcolsep}{9pt}
\begin{tabular}{llccccc}
\toprule
& & \multicolumn{4}{c}{Quark Contributions} & Total \\
\cmidrule(lr){3-6}
Current & Configuration
  & $\mu_u$ & $\mu_d$ & $\mu_s$ & $\mu_c$ & $\mu_{\mathrm{tot}}$ \\
\midrule
\multicolumn{7}{l}{\textit{Spin-$\tfrac{1}{2}$ states}} \\[3pt]
\multirow{2}{*}{$J_1(x)$}
  & $[ds][sc]\bar{c}$ & $-$      & $\phantom{-}0.02$  & $-0.04$ & $-1.46$ & $-1.48$ \\
  & $[us][sc]\bar{c}$ & $-0.04$  & $-$                & $-0.04$ & $-1.46$ & $-1.54$ \\[4pt]
\multirow{2}{*}{$J_2(x)$}
  & $[ds][sc]\bar{c}$ & $-$      & $-0.015$           & $\phantom{-}2.81$ & $\phantom{-}1.32$ & $\phantom{-}4.12$ \\
  & $[us][sc]\bar{c}$ & $\phantom{-}0.03$ & $-$       & $\phantom{-}2.81$ & $\phantom{-}1.32$ & $\phantom{-}4.16$ \\[4pt]
\multirow{2}{*}{$J_3(x)$}
  & $[ds][sc]\bar{c}$ & $-$      & $-0.04$            & $\phantom{-}5.66$ & $\phantom{-}0.00$ & $\phantom{-}5.62$ \\
  & $[us][sc]\bar{c}$ & $\phantom{-}0.08$ & $-$       & $\phantom{-}5.66$ & $\phantom{-}0.00$ & $\phantom{-}5.74$ \\[4pt]
\multirow{2}{*}{$J_4(x)$}
  & $[ds][sc]\bar{c}$ & $-$      & $\phantom{-}0.03$  & $-0.06$ & $-2.03$ & $-2.06$ \\
  & $[us][sc]\bar{c}$ & $-0.06$  & $-$                & $-0.06$ & $-2.03$ & $-2.15$ \\
\midrule
\multicolumn{7}{l}{\textit{Spin-$\tfrac{3}{2}$ states}} \\[3pt]
\multirow{2}{*}{$J_\mu^2(x)$}
  & $[ds][sc]\bar{c}$ & $-$      & $\phantom{-}0.02$  & $\phantom{-}0.05$ & $-0.49$ & $-0.43$ \\
  & $[us][sc]\bar{c}$ & $-0.04$  & $-$                & $\phantom{-}0.05$ & $-0.49$ & $-0.48$ \\[4pt]
\multirow{2}{*}{$J_\mu^3(x)$}
  & $[ds][sc]\bar{c}$ & $-$      & $\phantom{-}0.84$  & $-0.02$ & $-1.96$ & $-1.14$ \\
  & $[us][sc]\bar{c}$ & $-1.68$  & $-$                & $-0.02$ & $-1.96$ & $-3.66$ \\[4pt]
\multirow{2}{*}{$J_\mu^4(x)$}
  & $[ds][sc]\bar{c}$ & $-$      & $-0.04$            & $-0.39$ & $-3.82$ & $-4.25$ \\
  & $[us][sc]\bar{c}$ & $\phantom{-}0.08$ & $-$       & $-0.39$ & $-3.82$ & $-4.13$ \\
\bottomrule
\end{tabular}
\end{table}  

\begin{figure}[htp]
\centering
\includegraphics[width=0.85\textwidth]{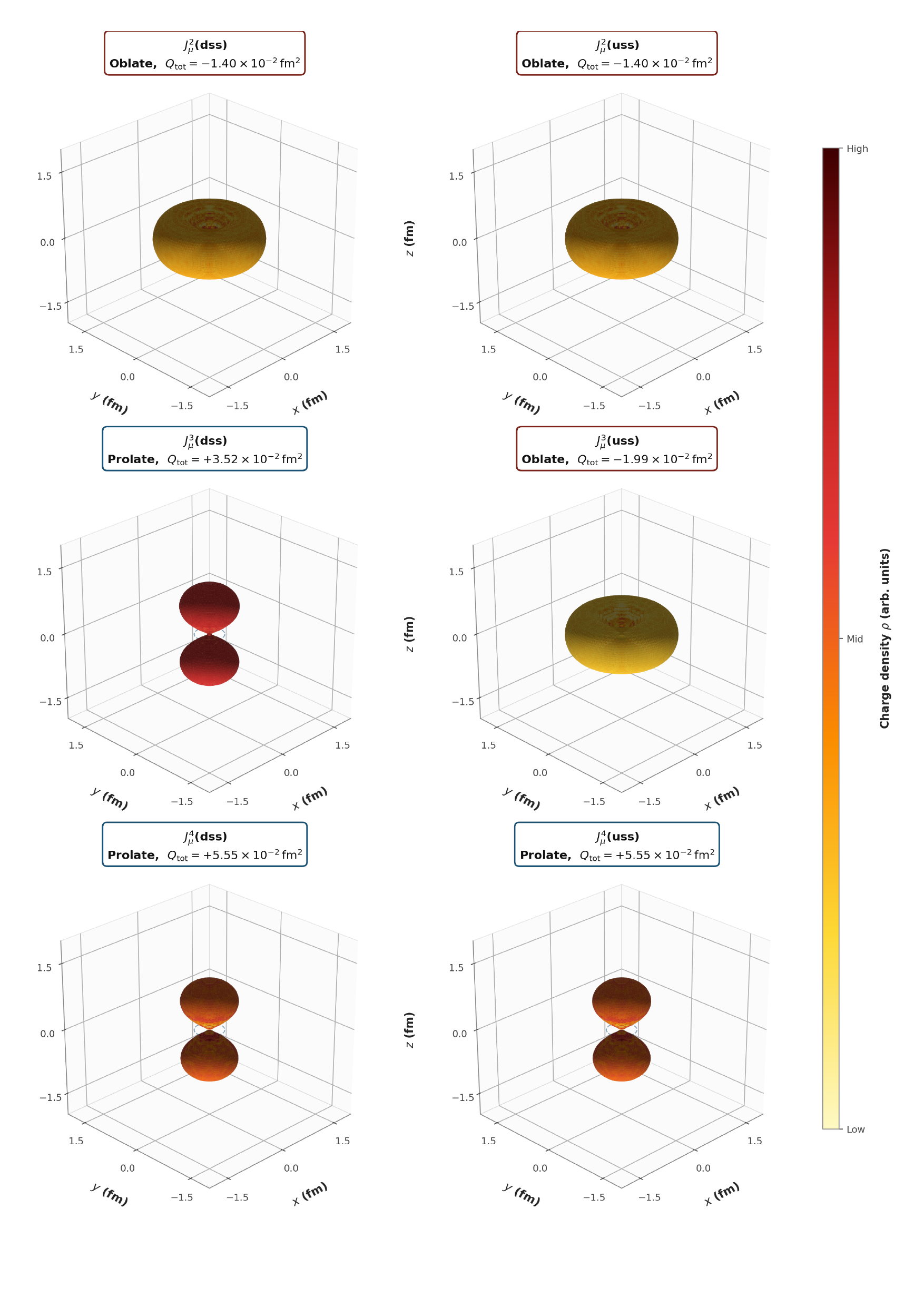}
\caption{
Three-dimensional charge density isosurfaces illustrating the electric quadrupole  deformation of $J_\mu^{2}$, $J_\mu^{3}$, and $J_\mu^{4}$ pentaquark states. Left and right columns correspond to $[ds][s c] \bar c$ and $[us][s c] \bar c$  quark configurations, respectively. The isosurface is extracted at 12\% of the maximum charge density using the marching cubes algorithm. The color map encodes the local charge density $\rho$ on the surface. Dashed curves mark the equatorial boundary. States with $Q_{\rm tot} > 0$ exhibit prolate deformation elongated along the $z$-axis; states with $Q_{\rm tot} < 0$ are oblately deformed in the $xy$-plane.
}
 \label{fig:quadrupole}
  \end{figure}
\begin{figure}[htp]
\centering
\includegraphics[width=1.0\textwidth]{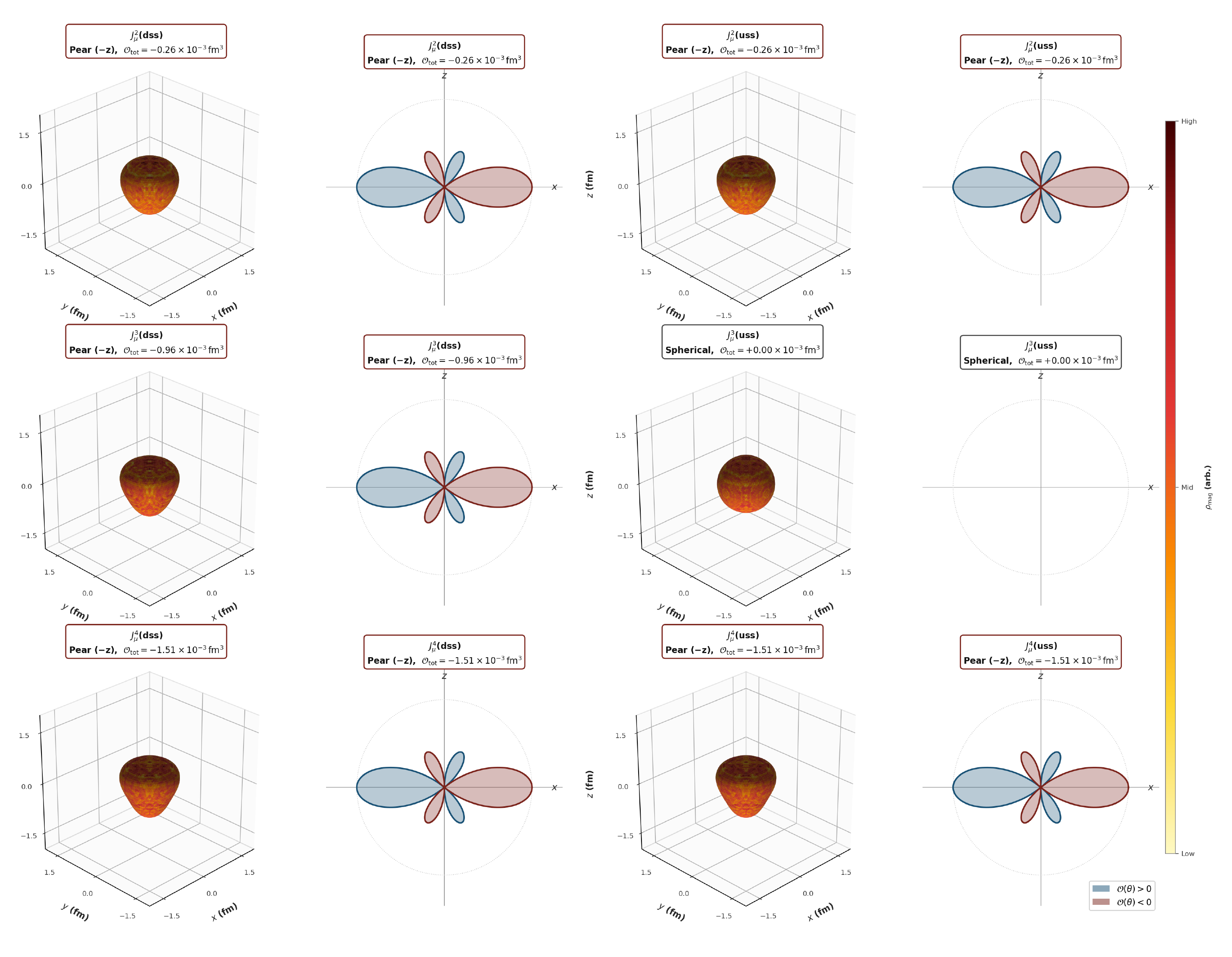}
\caption{
Three-dimensional isosurfaces of the magnetic octupole moment density $\rho_{\rm mag}(\mathbf{r}) \propto \sum_q \mathcal{O}_q \cdot r^3 \cdot P_3(\cos\theta) \cdot e^{-r^2/a^2}$ (left panels of each pair) and angular distributions $\mathcal{O}(\theta) = \sum_q \mathcal{O}_q \cdot P_3(\cos\theta)$ in the $xz$-plane (right panels of each pair) for the $J_\mu^{2}$, $J_\mu^{3}$, and $J_\mu^{4}$ pentaquark states. The first and second pairs in each row correspond to $[ds][sc]\bar{c}$ and $[us][sc]\bar{c}$ quark configurations, respectively. The isosurface is extracted at 30\% of the maximum density using the marching cubes algorithm with a Gaussian radial ansatz ($a = 0.7$ fm), and the surface color encodes the local magnetic moment density $\rho_{\rm mag}$. In the angular distribution panels, blue (red) regions indicate positive (negative) octupole contributions. The pear-shaped asymmetry along the $z$-axis arises from the odd-parity $P_3(\cos\theta)$ multipole term, which breaks the $z \to -z$ reflection symmetry. The $J_\mu^3$($[us][sc]\bar{c}$) state yields a nearly spherical isosurface and equal and opposite angular lobes due to exact cancellation between $\mathcal{O}_u = +0.64$ and $\mathcal{O}_c = -0.64 \times 10^{-3}$ fm$^3$ contributions. All octupole moment values are given in units of $10^{-3}$ fm$^3$.
}
\label{fig:octupole}
  \end{figure}

Moving beyond the magnetic dipole, the electric quadrupole and magnetic octupole  moments provide a unique and complementary window into the spatial geometry and current distribution of the pentaquark---information that the dipole moment alone cannot access. As shown in Table~\ref{tab:quark_higher} and visualized in Figs.~\ref{fig:quadrupole}--\ref{fig:octupole}, these higher moments exhibit a rich pattern that closely parallels the flavor hierarchy observed in the dipole sector, with the charm quark again playing the leading role. Yet they add entirely new dimensions to our understanding.

The electric quadrupole moment reveals the shape of the charge distribution. The $J_\mu^2(x)$ states are oblate ($\mathcal{Q} = -1.40 \times 10^{-2}$ fm²), exhibiting a modest, pancake-like flattening that is insensitive to light-quark flavor. The $J_\mu^4(x)$ states are strongly prolate ($\mathcal{Q} = 5.55 \times 10^{-2}$ fm²), displaying the largest deformation in our study---a pronounced elongation that correlates with their large negative dipole moments. Most strikingly, the $J_\mu^3(x)$ current presents a configuration-dependent geometry: the $dss$ configuration is prolate ($3.52 \times 10^{-2}$ fm²), while the $uss$ configuration is oblate ($-2.01 \times 10^{-2}$ fm²). This is a dramatic and unprecedented finding: simply replacing a $d$ quark with a $u$ quark---a substitution that leaves the total charge unchanged---can flip the entire geometric shape of the pentaquark. This sensitivity directly reflects the same structural mechanism identified in the dipole sector: the axial-vector diquark structure of $J^3_\mu(x)$ isolates the light quark, allowing the charge asymmetry $e_u/e_d = -2$ to propagate fully into the quadrupole
moment and flip its sign under $d\to u$ substitution.

The magnetic octupole moment provides a high-resolution fingerprint of internal current asymmetries. For $J_\mu^2(x)$, the octupole moment is tiny ($-0.26 \times 10^{-3}$ fm³), indicating a nearly symmetric current distribution. For $J_\mu^4(x)$, it is substantially larger ($-1.51 \times 10^{-3}$ fm³), correlating with the strong prolate deformation. Once again, $J_\mu^3(x)$ proves most intriguing: the $[us][sc]\bar c$ configuration yields an octupole moment consistent with zero, while its $[ds][sc]\bar c$ counterpart shows a clear negative value ($-0.96 \times 10^{-3}$ fm³). The octupole moment vanishes in the $[us][sc]\bar{c}$ configuration due to an exact cancellation between $\mathcal{O}_u = +0.64$ and $\mathcal{O}_c = -0.64 \times 10^{-3}~\mathrm{fm}^3$, as shown in Table~\ref{tab:quark_higher}. This cancellation is a direct consequence of the same structural mechanism: the axial-vector diquark position of the light quark in
$J^3_\mu(x)$ amplifies its contribution to a magnitude comparable to the charm quark contribution, producing exact cancellation at the octupole level.Rather than a mysterious dynamical symmetry, this is a quantitatively predictable outcome of the current's Dirac structure, making $J^3_\mu(x)$ a structurally transparent and uniquely sensitive probe of light-quark contributions to higher electromagnetic multipoles.

Taken together, these electromagnetic multipole moments form a hierarchical fingerprint of hadronic structure. The magnetic dipole identifies the spin-correlation pattern, with its sign distinguishing scalar from axial-vector diquark configurations. The electric quadrupole maps the geometric shape, with $J_\mu^3(x)$ revealing the astonishing possibility of shape-flipping through a mere light-quark substitution. The magnetic octupole fills in the fine print of current asymmetries, with $J_\mu^3(x)$ again defying expectations by exhibiting perfect octupolar symmetry in one flavor configuration.

Crucially, the strong current dependence observed here is not an artifact of this particular calculation, but a recurring theme in the study of exotic hadrons~\cite{Ozdem:2025jda, Wang:2016dzu, Ozdem:2024rch, Gao:2021hmv, Li:2024wxr, Li:2024jlq, Ozdem:2024rqx, Ozdem:2024txt, Mutuk:2024ach, Ozdem:2024dbq, Ozdem:2024lpk, Azizi:2023gzv}. This body of evidence establishes a general principle: electromagnetic observables are intrinsically more discriminating probes of internal quark correlations than masses alone. While spectroscopy identifies the existence of a state, electromagnetic moments reveal its inner architecture. 
We emphasize that the spread of predictions across interpolating currents reflects the dependence of the LCSR predictions on the assumed internal configuration of the pentaquark, and should not be interpreted as a numerical uncertainty of the calculation itself. 
For experimentalists, accessing the higher multipoles---through processes such as photo- and electroproduction with polarized beams and targets---would provide valuable complementary information about these states. The distinctive behavior of the $J_\mu^3(x)$ current offers a characteristic signature that could, in principle, help discriminate among competing configurations if precise measurements become feasible. For theorists, the strong configuration dependence observed here highlights the importance of systematic approaches that explore multiple interpolating currents. Any comprehensive model of doubly strange hidden-charm pentaquarks should ultimately account not only for their masses but also for the broader pattern of electromagnetic responses suggested by this analysis.


\begin{table}[htbp]
\centering
\caption{Quark-level contributions to the electric quadrupole moments
($\times10^{-2}$~fm$^2$, upper) and magnetic octupole moments
($\times10^{-3}$~fm$^3$, lower) of the $P_{\psi ss}^{\Lambda}$
pentaquarks.}
\label{tab:quark_higher}
\renewcommand{\arraystretch}{1.25}
\setlength{\tabcolsep}{9pt}
\begin{tabular}{llccccc}
\toprule
& & \multicolumn{4}{c}{Quark Contributions} & Total \\
\cmidrule(lr){3-6}
Current & Configuration
  & $\mathcal{Q}_u$ & $\mathcal{Q}_d$
  & $\mathcal{Q}_s$ & $\mathcal{Q}_c$
  & $\mathcal{Q}_{\mathrm{tot}}$ \\
\midrule
\multirow{2}{*}{$J_\mu^2(x)$}
  & $[ds][sc]\bar{c}$ & $-$      & ${\sim}0$ & $\phantom{-}0.01$ & $-1.41$ & $-1.40$ \\
  & $[us][sc]\bar{c}$ & ${\sim}0$& $-$       & $\phantom{-}0.01$ & $-1.41$ & $-1.40$ \\[4pt]
\multirow{2}{*}{$J_\mu^3(x)$}
  & $[ds][sc]\bar{c}$ & $-$      & $\phantom{-}1.84$ & $-0.01$ & $\phantom{-}1.69$ & $\phantom{-}3.52$ \\
  & $[us][sc]\bar{c}$ & $-3.68$  & $-$               & $-0.01$ & $\phantom{-}1.69$ & $-2.01$ \\[4pt]
\multirow{2}{*}{$J_\mu^4(x)$}
  & $[ds][sc]\bar{c}$ & $-$      & ${\sim}0$ & $\phantom{-}3.55$ & $\phantom{-}2.00$ & $\phantom{-}5.55$ \\
  & $[us][sc]\bar{c}$ & ${\sim}0$& $-$       & $\phantom{-}3.55$ & $\phantom{-}2.00$ & $\phantom{-}5.55$ \\
\midrule[0.4pt]
& & \multicolumn{4}{c}{Quark Contributions} & Total \\
\cmidrule(lr){3-6}
& &
  $\mathcal{O}_u$ & $\mathcal{O}_d$ &
  $\mathcal{O}_s$ & $\mathcal{O}_c$ &
  $\mathcal{O}_{\mathrm{tot}}$ \\[3pt]
\multirow{2}{*}{$J_\mu^2(x)$}
  & $[ds][sc]\bar{c}$ & $-$      & ${\sim}0$ & ${\sim}0$ & $-0.26$ & $-0.26$ \\
  & $[us][sc]\bar{c}$ & ${\sim}0$& $-$       & ${\sim}0$ & $-0.26$ & $-0.26$ \\[4pt]
\multirow{2}{*}{$J_\mu^3(x)$}
  & $[ds][sc]\bar{c}$ & $-$      & $-0.32$            & ${\sim}0$ & $-0.64$ & $-0.96$ \\
  & $[us][sc]\bar{c}$ & $\phantom{-}0.64$ & $-$       & ${\sim}0$ & $-0.64$ & $\phantom{-}0.00$ \\[4pt]
\multirow{2}{*}{$J_\mu^4(x)$}
  & $[ds][sc]\bar{c}$ & $-$      & ${\sim}0$ & $-0.69$ & $-0.82$ & $-1.51$ \\
  & $[us][sc]\bar{c}$ & ${\sim}0$& $-$       & $-0.69$ & $-0.82$ & $-1.51$ \\
\bottomrule
\end{tabular}
\end{table}


\subsection{Comparison with quark--model predictions}

To assess the structural implications of our LCSR results, it is instructive to compare them with quark–model calculations representing two distinct dynamical pictures: (i) a coupled–channel molecular framework
dominated by the $\Xi_c^{\prime,*}D_s^{(*)}$ components \cite{Zhu:2025abk}, and (ii) a compact diquark–diquark–antiquark model \cite{Mutuk:2024ach}. Because these approaches rely on fundamentally different assumptions regarding the spatial structure and dominant degrees of freedom, they provide complementary benchmarks for interpreting the LCSR predictions.

\subsubsection*{Coupled--channel molecular model}

The molecular calculation of \cite{Zhu:2025abk} yields narrow and
strictly positive magnetic dipole moments,
\begin{align}
\mu(1/2^-) &= [0.48, 0.63]~\mu_N, \\
\mu(3/2^-) &= [1.55, 1.74]~\mu_N,
\end{align}
reflecting the additive contributions of the constituent hadrons in a
spatially extended, weakly bound system. In sharp contrast, our LCSR
results span much wider intervals and include both positive and negative
values:
\begin{align}
\mu(1/2^-) = [-2.15,\,5.74]~\mu_N,\\
\mu(3/2^-) = [-4.25,\,-0.43]~\mu_N.
\end{align}
No LCSR current reproduces the molecular ranges, and the sign flip in the
$3/2^-$ sector constitutes a qualitative discrepancy. The uniformly positive molecular moments arise from the coherent addition of constituent hadron moments in a spatially extended system, whereas the sign-varying LCSR results reflect competing spin alignments in a compact quark–gluon cluster.    Furthermore, the
coupled–channel hierarchy $\mu(3/2^-)>\mu(1/2^-)$, characteristic of
enhanced spin alignment in shallow bound systems, is absent in our results.

\subsubsection*{Diquark--diquark--antiquark quark model}

A quark–model analysis based on compact diquark correlations
\cite{Mutuk:2024ach} predicts broader intervals than the molecular
approach. For the $J^{P}=\tfrac{1}{2}^{-}$ states, the quark--model analysis of
\cite{Mutuk:2024ach} predicts the following intervals for the magnetic
dipole moments:
\begin{itemize}
    \item $[us][sc]\bar c$ configuration:
    \begin{itemize}
        \item $8_{1f}$ representation: 
        $\mu = [-0.377, \, -0.485]~\mu_N$,
        \item $8_{2f}$ representation:
        $\mu = [ -0.377 , \, 1.617]~\mu_N$.
    \end{itemize}

    \item $[ds][sc]\bar c$ configuration:
    \begin{itemize}
        \item $8_{1f}$ representation: 
        $\mu =[ -0.377 , \,-0.646]~\mu_N$,
        \item $8_{2f}$ representation:
        $\mu = [-0.244 , \, -0.377]~\mu_N$.
    \end{itemize}
\end{itemize}

For the $J^{P}=\tfrac{3}{2}^{-}$ states, the corresponding predictions are:
\begin{itemize}
    \item $[us][sc]\bar c$ configuration:
    \begin{itemize}
        \item $8_{1f}$ representation: 
        $\mu = [-1.199 , \, -1.535]~\mu_N$,
        \item $8_{2f}$ representation:
        $\mu = 1.861~\mu_N$.
    \end{itemize}

    \item $[ds][sc]\bar c$ configuration:
    \begin{itemize}
        \item $8_{1f}$ representation: 
        $\mu = [-1.233 , \, -1.535]~\mu_N$,
        \item $8_{2f}$ representation:
        $\mu = -0.930~\mu_N$,
    \end{itemize}
\end{itemize}
where $8_{1f}$ and $8_{2f}$ denote different flavor-spin representations 
in the diquark basis (see \cite{Mutuk:2024ach} for details).

These quark--model values exhibit moderate variation in both sign and magnitude, 
consistent with compact diquark dynamics. However, they are systematically 
smaller in magnitude than the LCSR predictions, which span $\pm(4$--$6)\mu_N$. 
This discrepancy highlights the structural sensitivity of electromagnetic moments: 
while quark models typically employ wave functions optimized at the hadronic scale 
with simplified spin--spin interactions, the LCSR approach probes short--distance 
quark--gluon correlations directly through interpolating currents that encode 
different diquark spin structures ($C\gamma_5$ for scalar, $C\gamma_\mu$ for axial--vector). The larger moments in the LCSR analysis likely arise from more pronounced 
spin--orbital contributions and stronger spin--spin correlations at short distances, 
which are inherently incorporated via the OPE and photon 
DAs. Thus, the broader spread of LCSR predictions reflects not 
only theoretical uncertainty but also the range of possible microscopic configurations  accessible within the same quantum--number sector.

Taken together, the molecular, diquark--model, and LCSR predictions outline three clearly distinct structural regimes. The molecular approach produces narrow, strictly positive magnetic moments characteristic of spatially extended, weakly bound $\Xi_c^{\prime,*}D_s^{(*)}$ configurations. The compact diquark--diquark--antiquark model yields moderately broader ranges, including limited sign variation, reflecting confined quark correlations at hadronic length scales. In contrast, the LCSR results exhibit the widest spreads, large magnitudes, and systematic sign changes---particularly the uniformly negative values in the $3/2^{-}$ sector---arising from the sensitivity of interpolating currents to short--distance quark--gluon dynamics. Consequently, precise measurements of the electromagnetic moments of future  $P^{\Lambda}_{\psi ss}$ candidates will provide a definitive discriminator 
among these competing structural pictures, offering a direct window into 
the quark--gluon substructure of doubly strange hidden-charm pentaquarks.


\section{Summary and conclusions}
\label{sec:conclusions}

In this work, we have performed a systematic QCD light-cone sum rule analysis of the electromagnetic multipole moments of the doubly strange hidden-charm pentaquarks with $J^{P}=1/2^{-}$ and $3/2^{-}$. By employing several independent interpolating currents, we have explored the model dependence inherent in sum-rule predictions and mapped the range of electromagnetic responses compatible with different internal diquark–diquark–antiquark configurations.

\subsection*{Principal findings}

Our principal findings may be summarized as follows:

\begin{itemize}

    \item Strong current dependence of magnetic dipole moments: 
    The magnetic moments span wide intervals, $\mu(1/2^{-}) \in [-2.15,\,5.74]~\mu_{N}$ and $\mu(3/2^{-}) \in [-4.25,\,-0.43]~\mu_{N}$—a spread of nearly $10\,\mu_N$ that reflects the extreme sensitivity of the electromagnetic structure to the underlying diquark configuration and spin–orbital couplings. This variation provides a quantitative measure of structural uncertainty at the level of interpolating currents.

    \item Dominant role of the charm quark: 
    A decomposition into quark-level contributions reveals that the charm quark typically provides the largest input to the multipole moments, owing to its heavy mass and electric charge. The heavy-quark sector therefore largely governs the overall magnitude of the electromagnetic response.

    \item Configuration-dependent behavior of strange quarks: 
    Strange-quark contributions vary substantially across currents. Axial-vector diquark structures lead to sizable positive inputs, whereas scalar diquark configurations yield smaller and negative ones. This pattern highlights the strong correlation between the diquark spin structure and the resulting multipole moments.

    \item Hierarchy of higher multipoles: 
    The quadrupole and octupole moments follow the expected hierarchy $\mu \gg \mathcal{Q} \gg \mathcal{O}$ and exhibit both oblate and prolate charge distributions. Their magnitudes are consistent with typical hadronic-scale deformations, yet their pattern across currents—particularly the shape-flipping behavior of $J_\mu^3(x)$—offers unique geometric insights.

    \item Isospin symmetry and its exception:     All currents respect approximate isospin symmetry except $J_\mu^3(x)$, where a significant shift is observed between the $[ds][sc]\bar c$ and $[us][sc]\bar c$ configurations. This sensitivity is a direct and quantitatively transparent
consequence of the axial-vector diquark structure of $J^3_\mu(x)$, which decouples the light quark from spin averaging and allows the full charge ratio $e_u/e_d = -2$ to propagate into all three
multipole moments simultaneously. The ratio $\mu_u/\mu_d = -2.00$ observed in
Table~\ref{tab:quark_magnetic} confirms this mechanism exactly, and the same amplification propagates consistently into the quadrupole sign flip and the octupole cancellation of the $[us][sc]\bar{c}$ configuration.

\end{itemize}

\subsection*{Methodological insights}

Our analysis also offers several methodological observations relevant to future LCSR studies:

\begin{itemize}

    \item The variation across currents provides a natural uncertainty band associated with the unknown internal wave function, turning current dependence from a potential weakness into a useful diagnostic tool.

    \item Electromagnetic moments exhibit stronger structural discrimination than spectroscopic predictions: currents yielding similar masses can produce strikingly different multipole moments. This underscores the importance of including electromagnetic observables in future studies of exotic hadrons.

    \item The quark-level decomposition clarifies the dynamical origin of individual multipole moments and highlights the central role of heavy and strange quark subsystems, offering a template for interpreting future experimental measurements.

\end{itemize}

\subsection*{Experimental prospects}

The most promising experimental channels for observing $P_{\psi ss}^{\Lambda}$
candidates are weak decays of bottom hadrons, in close analogy with the discovery channels of the strange pentaquarks $P_{\psi s}^{\Lambda}(4338)$ and $P_{\psi s}^{\Lambda}(4459)$ in $\Xi_b^- \to J/\psi \Lambda K^-$. The natural extensions to the $S=-2$ sector include $\Lambda_b^0 \to J/\psi \Xi^- K^+$ and
$\Xi_b^0 \to J/\psi \Xi^- \pi^+$, where the $J/\psi \Xi$
invariant-mass distribution would directly probe the existence of doubly strange hidden-charm pentaquarks. The CMS collaboration reported the first observation of $\Lambda_b^0 \to J/\psi \Xi^- K^+$ in 2024~\cite{CMS:2024vnm}, followed by a more precise LHCb measurement
that also provided the first observation of $\Xi_b^0 \to J/\psi \Xi^- \pi^+$ \cite{LHCb:2025lhk}. Theoretical studies based on coupled-channel unitary approaches~\cite{Oset:2024fbk, Roca:2025zyi} have identified $\Omega_b^- \to K^- \eta_c \Xi^0$ as an additional
promising channel, with the $\Omega_b^-$ decays expected to yield particularly clean resonant signals. These initial datasets already provide a basis on which $P_{\psi ss}^{\Lambda}$ signals could be sought, and the increased $b$-hadron statistics anticipated at LHCb Upgrade II and at Belle II would substantially enhance the
discovery potential for these states.

Once such candidates are identified, the experimental determination
of their electromagnetic moments presents a further challenge.
Radiative processes of the type
$\gamma^{(*)}\,p \to P_{\psi ss}^{\Lambda} \to J/\psi\,\Lambda\,K\,\gamma$
could in principle provide sensitivity to the electromagnetic
structure of the $P_{\psi ss}^{\Lambda}$ states through the angular
distribution of the emitted photon and the associated radiative
partial width. 
These observables encode the electromagnetic multipole form factors,
including $G_M$ and $G_Q$, whose values at $q^2=0$ define the magnetic
dipole and electric quadrupole moments via
$\mu_{P_{\psi ss}^{\Lambda}} = e\,G_M(0)/(2m_{P_{\psi ss}^{\Lambda}})$
and $\mathcal{Q}_{P_{\psi ss}^{\Lambda}} = e\,G_Q(0)/m_{P_{\psi ss}^{\Lambda}}^2$.
The branching fractions associated with such radiative transitions are
expected to be small, and the short lifetime of the $P_{\psi ss}^{\Lambda}$
states introduces additional experimental challenges.
Nevertheless, analogous radiative strategies have been successfully
pursued for the $\Delta^+(1232)$ resonance, where electromagnetic
multipole moments were constrained through exclusive photoproduction
and electroproduction measurements~\cite{Pascalutsa:2004je,
Pascalutsa:2005vq, Pascalutsa:2007wb}.
Searches for $J/\psi\,\Lambda\,K$ final states accompanied by an
isolated photon, or targeted reanalyses of existing $b$-hadron decay
data at LHCb and Belle~II, may therefore offer a path toward
constraining the electromagnetic structure of these states in the
longer term.

On the theoretical side, lattice QCD provides a complementary and
potentially more immediate route.
The methodology for computing electromagnetic form factors of heavy
baryons on the lattice is well established~\cite{Can:2013zpa,
Can:2013tna}, and its recent extension to multiquark
systems~\cite{Vujmilovic:2025czt} suggests that analogous calculations
for hidden-charm pentaquarks are within reach.
A lattice determination of the magnetic dipole moment of the
$P_{\psi ss}^{\Lambda}$ states would provide an ab initio benchmark
that is independent of the choice of interpolating current and could
therefore directly test the model dependence identified in the present
analysis.
The combination of such lattice inputs with future experimental
measurements would offer the most complete picture of the
electromagnetic structure of doubly strange hidden-charm pentaquarks.

\subsection*{Concluding remarks}

Overall, this work demonstrates that electromagnetic multipole moments constitute uniquely sensitive probes of exotic hadron structure. The broad range of predictions associated with different interpolating currents maps out a rich landscape of possible internal configurations and emphasizes the need for complementary theoretical and experimental investigations. As future facilities continue to explore the heavy-flavor sector with increasing precision, the multipole moments calculated here are expected to provide valuable guidance in determining whether doubly strange hidden-charm pentaquarks are best interpreted as compact multiquark states, hadronic molecules, or more complex configurations. In this endeavor, the electromagnetic moments serve not merely as additional observables, but as essential fingerprints that can reveal the inner architecture of these enigmatic states.


  \newpage

  \begin{widetext} 

 \section*{Appendix: Analytical results for the $J_1$ current}
\label{app:J1_results}

This appendix provides the complete analytical expressions obtained for the magnetic dipole moment of the doubly strange hidden-charm pentaquark state described by the interpolating current $J_1$, defined in Eq.~(\ref{eq:current_1_12}). The result follows from matching the QCD representation of the correlation function to its hadronic counterpart, applying the Borel transformation, and performing the continuum subtraction according to the standard light-cone sum rule procedure outlined in Sec.~\ref{sec:formalism}.

The magnetic dipole moment $\mu_{P^{\Lambda}_{\psi ss}}(J_1)$ is extracted from the sum rule
\[
\mu_{P^{\Lambda}_{\psi ss}}(J_1) = \frac{\Pi_1(M^2, s_0)}{\lambda_{P^{\Lambda}_{\psi ss}}^2(J_1)}\, e^{m^2(J_1)/M^2},
\]
where the spectral density $\Pi_1(M^2, s_0)$ receives contributions from both perturbative photon emission and non-perturbative photon distribution amplitudes. After carrying out the operator product expansion up to dimension-7 and including contributions from twist-2, twist-3, and twist-4 photon DAs, we obtain
 \begin{align}
\Pi_1(M^2, s_0)&= \frac{e_c}{2^{25}\times 3^4 \times 5^3 \times 7 ^2 \pi^7} \Big[10780 m_c m_s I[0, 6] - 1647 I[0, 7]  
 \Big]
 \nonumber\\
 &- \frac{ \langle g_s^2 G^2\rangle \langle \bar q q \rangle }{2^{24}\times 3^4 \times 5  \pi^5}  \Big[ 23 e_q m_s  I_4[\mathcal S] I[0, 3]\Big]
   \nonumber\\
 &+ \frac{e_s \, m_c \langle g_s^2 G^2 \rangle \langle \bar s s \rangle }{2^{24} \times 3^6 \times 5 \pi^5}  
\Big[ 40 \mathbb A[u_0] I[0, 3] + 207 I_3[\mathcal S] I[0, 3] \Big]\nonumber\\
 &+ \frac{11 \langle g_s^2 G^2\rangle f_{3\gamma}}{2^{29}\times 3^6 \times 5  \pi^5}
 \Big[ 9 e_s I_1[\mathcal V] + 4 e_q I_2[\mathcal V]) I[0, 4] - 
 32 (44 (e_q - e_s) m_c m_s I[0, 3] + (-9 e_q + 17 e_s) I[0, 4]) \psi^a[
   u_0]
 \Big]
  \nonumber\\
   &- \frac{e_s \, m_c \langle g_s^2 G^2 \rangle \langle \bar s s \rangle \,\chi }{2^{20} \times 3^6 \times 5 \pi^5}  
  I[0, 4] \varphi_\gamma[u_0]\nonumber\\
 &- \frac{e_q \,m_s\langle \bar q q \rangle}{2^{22}\times 3 \times 5  \pi^5} I_4[\mathcal S] I[0, 5]
\nonumber\\
 &+\frac{e_s \,m_c\langle \bar ss \rangle}{2^{25}\times 3 \times 5^2  \pi^5}  I_3[\mathcal S] I[0, 5]
\nonumber\\
 &+ \frac{f_{3\gamma}}{2^{26}\times 3^2 \times 5  \pi^5}\Big[2 e_s I_1[\mathcal V] (12 m_c m_s I[0, 5] - 5 I[0, 6]) + 
 e_q I_2[\mathcal V] (24 m_c m_s I[0, 5] - 5 I[0, 6])\Big],\label{F1sonuc}
 \end{align}
 \noindent  where the functions $\mathrm{I}[n,m]$ and $ I_i[\mathcal{F}]$  are listed as      
\begin{align}
 \mathrm{I}[k,l]&= \int_{\mathcal M}^{s_0} ds~ e^{-s/M^2}~
 s^k\,(s-\mathcal M)^l,\nonumber\\ 
I_1[\mathcal{F}]&=\int D_{\alpha_i} \int_0^1 dv~ \mathcal{F}(\alpha_{\bar q},\alpha_q,\alpha_g)  \delta'(\alpha_ q +\bar v \alpha_g-u_0),\nonumber\\
  I_2[\mathcal{F}]&=\int D_{\alpha_i} \int_0^1 dv~ \mathcal{F}(\alpha_{\bar q},\alpha_q,\alpha_g)  \delta'(\alpha_{\bar q}+ v \alpha_g-u_0),
 \nonumber\\
     I_3[\mathcal{F}]&=\int D_{\alpha_i} \int_0^1 dv~ \mathcal{F}(\alpha_{\bar q},\alpha_q,\alpha_g)
 \delta(\alpha_ q +\bar v \alpha_g-u_0),\nonumber\\
   I_4[\mathcal{F}]&=\int D_{\alpha_i} \int_0^1 dv~ \mathcal{F}(\alpha_{\bar q},\alpha_q,\alpha_g)
 \delta(\alpha_{\bar q}+ v \alpha_g-u_0),
 \end{align}
\noindent where $\mathcal M = (2m_c + 2m_s)^2$, and $\mathcal{F}$ denotes the photon DAs.

The standard double Borel transformation procedure is implemented to suppress contributions from higher states and enhance the ground-state signal in the sum rules. For the hadronic side, the transformation acts on the double pole structure as
\begin{equation}
\label{eq:borel_hadronic}
\mathcal{B}_{M_1^2, M_2^2} \!\left\{ \frac{1}{[p^2 - m_i^2][(p+q)^2 - m_f^2]} \right\} 
= e^{-m_i^2/M_1^2 - m_f^2/M_2^2},
\end{equation}
where $M_1^2$ and $M_2^2$ are the Borel parameters associated with the initial and final hadron momenta $-p^2$ and $-(p+q)^2$, respectively. On the QCD side, after performing the Fourier transformation to momentum space, the typical denominator that appears in light-cone expansions has the form $(m^2 - \bar{u}p^2 - u(p+q)^2)^\alpha$, which transforms as
\begin{equation}
\label{eq:borel_qcd}
\mathcal{B}_{M_1^2, M_2^2} \!\left\{ \frac{1}{\big(m^2 - \bar{u}p^2 - u(p+q)^2\big)^\alpha} \right\} 
= (M^2)^{2-\alpha} \, \delta(u - u_0) \, e^{-m^2/M^2},
\end{equation}
with the effective Borel mass $M^2$ and momentum fraction $u_0$ given by
\begin{equation}
\label{eq:borel_params}
M^2 = \frac{M_1^2 M_2^2}{M_1^2 + M_2^2}, \qquad
u_0 = \frac{M_1^2}{M_1^2 + M_2^2}.
\end{equation}

In the present analysis, the initial and final states correspond to the same pentaquark, implying $m_i = m_f = m_{P^{\Lambda}_{\psi ss}}$. The most natural choice is therefore to take $M_1^2 = M_2^2$, which reduces the two Borel parameters to a single one, $M^2$, through Eq.~(\ref{eq:borel_params}). Setting $M_1^2 = M_2^2 = 2M^2$ yields the particularly simple relations

\begin{equation}
M^2 = \frac{M_1^2}{2}, \qquad u_0 = \frac{1}{2}.
\end{equation}

This symmetric assignment ensures that the exponential suppression acts equally on both momentum channels, maximizing the ground-state contribution while minimizing contamination from the continuum. Physically, $M^2$ represents the characteristic virtuality scale of the heavy–light quark system probed by the correlation function. The choice $u_0 = 1/2$ reflects the fact that, for elastic scattering of identical hadrons, the photon couples symmetrically to the initial and final states, carrying on average half of the momentum transfer.  
Adopting a single Borel parameter $M^2$ not only simplifies the numerical analysis but also improves the stability of the sum rules, because it eliminates artificial imbalances that could arise from treating $M_1^2$ and $M_2^2$ independently when the underlying states are identical. This approach is well established in studies of elastic form factors of heavy hadrons within the LCSR framework and has been shown to provide reliable and stable predictions~\cite{Ozdem:2024dbq}.

\end{widetext}

\bibliographystyle{elsarticle-num}
\bibliography{PcssMMv2.bib}

@article{Belle:2003nnu,
    author = "Choi, S. K. and others",
    collaboration = "Belle",
    title = "{Observation of a narrow charmonium-like state in exclusive $B^\pm \to K^\pm \pi^+ \pi^- J/\psi$ decays}",
    eprint = "hep-ex/0309032",
    archivePrefix = "arXiv",
    doi = "10.1103/PhysRevLett.91.262001",
    journal = "Phys. Rev. Lett.",
    volume = "91",
    pages = "262001",
    year = "2003"
}

@article{Esposito:2014rxa,
    author = "Esposito, Angelo and Guerrieri, Andrea L. and Piccinini, Fulvio and Pilloni, Alessandro and Polosa, Antonio D.",
    title = "{Four-Quark Hadrons: an Updated Review}",
    eprint = "1411.5997",
    archivePrefix = "arXiv",
    primaryClass = "hep-ph",
    doi = "10.1142/S0217751X15300021",
    journal = "Int. J. Mod. Phys. A",
    volume = "30",
    pages = "1530002",
    year = "2015"
}

@article{Esposito:2016noz,
    author = "Esposito, A. and Pilloni, A. and Polosa, A. D.",
    title = "{Multiquark Resonances}",
    eprint = "1611.07920",
    archivePrefix = "arXiv",
    primaryClass = "hep-ph",
    reportNumber = "JLAB-THY-16-2301",
    doi = "10.1016/j.physrep.2016.11.002",
    journal = "Phys. Rept.",
    volume = "668",
    pages = "1--97",
    year = "2017"
}

@article{Olsen:2017bmm,
    author = "Olsen, Stephen Lars and Skwarnicki, Tomasz and Zieminska, Daria",
    title = "{Nonstandard heavy mesons and baryons: Experimental evidence}",
    eprint = "1708.04012",
    archivePrefix = "arXiv",
    primaryClass = "hep-ph",
    doi = "10.1103/RevModPhys.90.015003",
    journal = "Rev. Mod. Phys.",
    volume = "90",
    number = "1",
    pages = "015003",
    year = "2018"
}

@article{Lebed:2016hpi,
    author = "Lebed, Richard F. and Mitchell, Ryan E. and Swanson, Eric S.",
    title = "{Heavy-Quark QCD Exotica}",
    eprint = "1610.04528",
    archivePrefix = "arXiv",
    primaryClass = "hep-ph",
    doi = "10.1016/j.ppnp.2016.11.003",
    journal = "Prog. Part. Nucl. Phys.",
    volume = "93",
    pages = "143--194",
    year = "2017"
}

@article{Nielsen:2009uh,
    author = "Nielsen, Marina and Navarra, Fernando S. and Lee, Su Houng",
    title = "{New Charmonium States in QCD Sum Rules: A Concise Review}",
    eprint = "0911.1958",
    archivePrefix = "arXiv",
    primaryClass = "hep-ph",
    doi = "10.1016/j.physrep.2010.07.005",
    journal = "Phys. Rept.",
    volume = "497",
    pages = "41--83",
    year = "2010"
}

@article{Brambilla:2019esw,
    author = "Brambilla, Nora and Eidelman, Simon and Hanhart, Christoph and Nefediev, Alexey and Shen, Cheng-Ping and Thomas, Christopher E. and Vairo, Antonio and Yuan, Chang-Zheng",
    title = "{The $XYZ$ states: experimental and theoretical status and perspectives}",
    eprint = "1907.07583",
    archivePrefix = "arXiv",
    primaryClass = "hep-ex",
    reportNumber = "TUM-EFT 125/19",
    doi = "10.1016/j.physrep.2020.05.001",
    journal = "Phys. Rept.",
    volume = "873",
    pages = "1--154",
    year = "2020"
}

@article{Agaev:2020zad,
    author = "Agaev, Shahin and Azizi, Kazem and Sundu, Hayriye",
    title = "{Four-quark exotic mesons}",
    eprint = "2004.12079",
    archivePrefix = "arXiv",
    primaryClass = "hep-ph",
    doi = "10.3906/fiz-2003-15",
    journal = "Turk. J. Phys.",
    volume = "44",
    number = "2",
    pages = "95--173",
    year = "2020"
}

@article{Chen:2016qju,
    author = "Chen, Hua-Xing and Chen, Wei and Liu, Xiang and Zhu, Shi-Lin",
    title = "{The hidden-charm pentaquark and tetraquark states}",
    eprint = "1601.02092",
    archivePrefix = "arXiv",
    primaryClass = "hep-ph",
    doi = "10.1016/j.physrep.2016.05.004",
    journal = "Phys. Rept.",
    volume = "639",
    pages = "1--121",
    year = "2016"
}

@article{Ali:2017jda,
    author = {Ali, Ahmed and Lange, Jens S{\"o}ren and Stone, Sheldon},
    title = "{Exotics: Heavy Pentaquarks and Tetraquarks}",
    eprint = "1706.00610",
    archivePrefix = "arXiv",
    primaryClass = "hep-ph",
    reportNumber = "DESY-17-071",
    doi = "10.1016/j.ppnp.2017.08.003",
    journal = "Prog. Part. Nucl. Phys.",
    volume = "97",
    pages = "123--198",
    year = "2017"
}

@article{Guo:2017jvc,
    author = "Guo, Feng-Kun and Hanhart, Christoph and Mei{\ss}ner, Ulf-G. and Wang, Qian and Zhao, Qiang and Zou, Bing-Song",
    title = "{Hadronic molecules}",
    eprint = "1705.00141",
    archivePrefix = "arXiv",
    primaryClass = "hep-ph",
    doi = "10.1103/RevModPhys.90.015004",
    journal = "Rev. Mod. Phys.",
    volume = "90",
    number = "1",
    pages = "015004",
    year = "2018",
    note = "[Erratum: Rev.Mod.Phys. 94, 029901 (2022)]"
}

@article{Liu:2019zoy,
    author = "Liu, Yan-Rui and Chen, Hua-Xing and Chen, Wei and Liu, Xiang and Zhu, Shi-Lin",
    title = "{Pentaquark and Tetraquark states}",
    eprint = "1903.11976",
    archivePrefix = "arXiv",
    primaryClass = "hep-ph",
    doi = "10.1016/j.ppnp.2019.04.003",
    journal = "Prog. Part. Nucl. Phys.",
    volume = "107",
    pages = "237--320",
    year = "2019"
}

@article{Yang:2020atz,
    author = "Yang, Gang and Ping, Jialun and Segovia, Jorge",
    title = "{Tetra- and penta-quark structures in the constituent quark model}",
    eprint = "2009.00238",
    archivePrefix = "arXiv",
    primaryClass = "hep-ph",
    doi = "10.3390/sym12111869",
    journal = "Symmetry",
    volume = "12",
    number = "11",
    pages = "1869",
    year = "2020"
}

@article{Dong:2021juy,
    author = "Dong, Xiang-Kun and Guo, Feng-Kun and Zou, Bing-Song",
    title = "{A survey of heavy-antiheavy hadronic molecules}",
    eprint = "2101.01021",
    archivePrefix = "arXiv",
    primaryClass = "hep-ph",
    doi = "10.13725/j.cnki.pip.2021.02.001",
    journal = "Progr. Phys.",
    volume = "41",
    pages = "65--93",
    year = "2021"
}

@article{Dong:2021bvy,
    author = "Dong, Xiang-Kun and Guo, Feng-Kun and Zou, Bing-Song",
    title = "{A survey of heavy{\textendash}heavy hadronic molecules}",
    eprint = "2108.02673",
    archivePrefix = "arXiv",
    primaryClass = "hep-ph",
    doi = "10.1088/1572-9494/ac27a2",
    journal = "Commun. Theor. Phys.",
    volume = "73",
    number = "12",
    pages = "125201",
    year = "2021"
}

@article{Chen:2022asf,
    author = "Chen, Hua-Xing and Chen, Wei and Liu, Xiang and Liu, Yan-Rui and Zhu, Shi-Lin",
    title = "{An updated review of the new hadron states}",
    eprint = "2204.02649",
    archivePrefix = "arXiv",
    primaryClass = "hep-ph",
    doi = "10.1088/1361-6633/aca3b6",
    journal = "Rept. Prog. Phys.",
    volume = "86",
    number = "2",
    pages = "026201",
    year = "2023"
}

@article{Meng:2022ozq,
    author = "Meng, Lu and Wang, Bo and Wang, Guang-Juan and Zhu, Shi-Lin",
    title = "{Chiral perturbation theory for heavy hadrons and chiral effective field theory for heavy hadronic molecules}",
    eprint = "2204.08716",
    archivePrefix = "arXiv",
    primaryClass = "hep-ph",
    doi = "10.1016/j.physrep.2023.04.003",
    journal = "Phys. Rept.",
    volume = "1019",
    pages = "1--149",
    year = "2023"
}

@article{LHCb:2015yax,
    author = "Aaij, Roel and others",
    collaboration = "LHCb",
    title = "{Observation of $J/\psi p$ Resonances Consistent with Pentaquark States in $\Lambda_b^0 \to J/\psi K^- p$ Decays}",
    eprint = "1507.03414",
    archivePrefix = "arXiv",
    primaryClass = "hep-ex",
    reportNumber = "CERN-PH-EP-2015-153, LHCB-PAPER-2015-029",
    doi = "10.1103/PhysRevLett.115.072001",
    journal = "Phys. Rev. Lett.",
    volume = "115",
    pages = "072001",
    year = "2015"
}

@article{LHCb:2019kea,
    author = "Aaij, Roel and others",
    collaboration = "LHCb",
    title = "{Observation of a narrow pentaquark state, $P_c(4312)^+$, and of two-peak structure of the $P_c(4450)^+$}",
    eprint = "1904.03947",
    archivePrefix = "arXiv",
    primaryClass = "hep-ex",
    reportNumber = "LHCb-PAPER-2019-014 CERN-EP-2019-058",
    doi = "10.1103/PhysRevLett.122.222001",
    journal = "Phys. Rev. Lett.",
    volume = "122",
    number = "22",
    pages = "222001",
    year = "2019"
}

@article{LHCb:2020jpq,
    author = "Aaij, Roel and others",
    collaboration = "LHCb",
    title = "{Evidence of a $J/\psi\Lambda$ structure and observation of excited $\Xi^-$ states in the $\Xi^-_b \to J/\psi\Lambda K^-$ decay}",
    eprint = "2012.10380",
    archivePrefix = "arXiv",
    primaryClass = "hep-ex",
    reportNumber = "LHCb-PAPER-2020-039, CERN-EP-2020-233",
    doi = "10.1016/j.scib.2021.02.030",
    journal = "Sci. Bull.",
    volume = "66",
    pages = "1278--1287",
    year = "2021"
}

@article{LHCb:2022ogu,
    author = "Aaij, R. and others",
    collaboration = "LHCb",
    title = "{Observation of a J/{\ensuremath{\psi}}{\ensuremath{\Lambda}} Resonance Consistent with a Strange Pentaquark Candidate in B-{\textrightarrow}J/{\ensuremath{\psi}}{\ensuremath{\Lambda}}p{\textasciimacron} Decays}",
    eprint = "2210.10346",
    archivePrefix = "arXiv",
    primaryClass = "hep-ex",
    reportNumber = "CERN-EP-2022-198, LHCb-PAPER-2022-031",
    doi = "10.1103/PhysRevLett.131.031901",
    journal = "Phys. Rev. Lett.",
    volume = "131",
    number = "3",
    pages = "031901",
    year = "2023"
}

@article{Belle:2025pey,
    author = "Adachi, I. and others",
    collaboration = "Belle, Belle-II",
    title = "{Search for Pcs(4459) and Pcs(4338) in Upsilon(1S,2S) inclusive decays at Belle}",
    eprint = "2502.09951",
    archivePrefix = "arXiv",
    primaryClass = "hep-ex",
    reportNumber = "Belle II Preprint 2025-002, KEK Preprint 2024-50",
    doi = "10.1103/pf8m-6j69",
    journal = "Phys. Rev. Lett.",
    volume = "135",
    number = "4",
    pages = "041901",
    year = "2025"
}

@article{Wang:2016dzu,
    author = "Wang, Guang-Juan and Chen, Rui and Ma, Li and Liu, Xiang and Zhu, Shi-Lin",
    title = "{Magnetic moments of the hidden-charm pentaquark states}",
    eprint = "1605.01337",
    archivePrefix = "arXiv",
    primaryClass = "hep-ph",
    doi = "10.1103/PhysRevD.94.094018",
    journal = "Phys. Rev. D",
    volume = "94",
    number = "9",
    pages = "094018",
    year = "2016"
}

@article{Ozdem:2018qeh,
    author = {{\"O}zdem, U. and Azizi, K.},
    title = "{Electromagnetic multipole moments of the $P_c^+(4380)$ pentaquark in light-cone QCD}",
    eprint = "1803.06831",
    archivePrefix = "arXiv",
    primaryClass = "hep-ph",
    doi = "10.1140/epjc/s10052-018-5873-2",
    journal = "Eur. Phys. J. C",
    volume = "78",
    number = "5",
    pages = "379",
    year = "2018"
}

@article{Ortiz-Pacheco:2018ccl,
    author = "Ortiz-Pacheco, Emmanuel and Bijker, Roelof and Fern{\'a}ndez-Ram{\'\i}rez, C{\'e}sar",
    title = "{Hidden charm pentaquarks: mass spectrum, magnetic moments, and photocouplings}",
    eprint = "1808.10512",
    archivePrefix = "arXiv",
    primaryClass = "nucl-th",
    doi = "10.1088/1361-6471/ab096d",
    journal = "J. Phys. G",
    volume = "46",
    number = "6",
    pages = "065104",
    year = "2019"
}

@article{Xu:2020flp,
    author = "Xu, Yong-Jiang and Liu, Yong-Lu and Huang, Ming-Qiu",
    title = "{The magnetic moment of $P_{c}(4312)$ as a $\bar{D}\Sigma _{c}$ molecular state}",
    eprint = "2008.07937",
    archivePrefix = "arXiv",
    primaryClass = "hep-ph",
    doi = "10.1140/epjc/s10052-021-09211-8",
    journal = "Eur. Phys. J. C",
    volume = "81",
    number = "5",
    pages = "421",
    year = "2021"
}

@article{Ozdem:2021btf,
    author = {{\"O}zdem, Ula{\c{s}}},
    title = "{Electromagnetic properties of the $P_c$ (4312) pentaquark state}",
    doi = "10.1088/1674-1137/abd01c",
    journal = "Chin. Phys. C",
    volume = "45",
    number = "2",
    pages = "023119",
    year = "2021"
}

@article{Ozdem:2021ugy,
    author = {{\"O}zdem, Ula{\c{s}}},
    title = "{Magnetic dipole moments of the hidden-charm pentaquark states: $P_c(4440)$, $P_c(4457)$ and $P_{cs}(4459)$}",
    eprint = "2102.01996",
    archivePrefix = "arXiv",
    primaryClass = "hep-ph",
    doi = "10.1140/epjc/s10052-021-09070-3",
    journal = "Eur. Phys. J. C",
    volume = "81",
    number = "4",
    pages = "277",
    year = "2021"
}

@article{Li:2021ryu,
    author = "Li, Ming-Wei and Liu, Zhan-Wei and Sun, Zhi-Feng and Chen, Rui",
    title = "{Magnetic moments and transition magnetic moments of Pc and Pcs states}",
    eprint = "2106.15053",
    archivePrefix = "arXiv",
    primaryClass = "hep-ph",
    doi = "10.1103/PhysRevD.104.054016",
    journal = "Phys. Rev. D",
    volume = "104",
    number = "5",
    pages = "054016",
    year = "2021"
}

@article{Ozdem:2023htj,
    author = {{\"O}zdem, Ula{\c{s}}},
    title = "{Electromagnetic properties of D{\textasciimacron}({\textasteriskcentered}){\ensuremath{\Xi}}c', D{\textasciimacron}({\textasteriskcentered}){\ensuremath{\Lambda}}c, D{\textasciimacron}s({\textasteriskcentered}){\ensuremath{\Lambda}}c and D{\textasciimacron}s({\textasteriskcentered}){\ensuremath{\Xi}}c pentaquarks}",
    eprint = "2303.10649",
    archivePrefix = "arXiv",
    primaryClass = "hep-ph",
    doi = "10.1016/j.physletb.2023.138267",
    journal = "Phys. Lett. B",
    volume = "846",
    pages = "138267",
    year = "2023"
}

@article{Wang:2023iox,
    author = "Wang, Fu-Lai and Liu, Xiang",
    title = "{Higher molecular P{\ensuremath{\psi}}s{\ensuremath{\Lambda}}/{\ensuremath{\Sigma}} pentaquarks arising from the {\ensuremath{\Xi}}c(',*)D{\textasciimacron}1/{\ensuremath{\Xi}}c(',*)D{\textasciimacron}2* interactions}",
    eprint = "2307.08276",
    archivePrefix = "arXiv",
    primaryClass = "hep-ph",
    doi = "10.1103/PhysRevD.108.054028",
    journal = "Phys. Rev. D",
    volume = "108",
    number = "5",
    pages = "054028",
    year = "2023"
}

@article{Ozdem:2022kei,
    author = {{\"O}zdem, Ula{\c{s}}},
    title = "{Investigation of magnetic moment of Pcs(4338) and Pcs(4459) pentaquark states}",
    eprint = "2208.07684",
    archivePrefix = "arXiv",
    primaryClass = "hep-ph",
    doi = "10.1016/j.physletb.2022.137635",
    journal = "Phys. Lett. B",
    volume = "836",
    pages = "137635",
    year = "2023"
}

@article{Gao:2021hmv,
    author = "Gao, Feng and Li, Hao-Song",
    title = "{Magnetic moments of hidden-charm strange pentaquark states*}",
    eprint = "2112.01823",
    archivePrefix = "arXiv",
    primaryClass = "hep-ph",
    doi = "10.1088/1674-1137/ac8651",
    journal = "Chin. Phys. C",
    volume = "46",
    number = "12",
    pages = "123111",
    year = "2022"
}

@article{Guo:2023fih,
    author = "Guo, Fei and Li, Hao-Song",
    title = "{Analysis of the hidden-charm pentaquark states based on magnetic moment and transition magnetic moment}",
    eprint = "2304.10981",
    archivePrefix = "arXiv",
    primaryClass = "hep-ph",
    doi = "10.1140/epjc/s10052-024-12699-5",
    journal = "Eur. Phys. J. C",
    volume = "84",
    number = "4",
    pages = "392",
    year = "2024"
}

@article{Ozdem:2022iqk,
    author = {{\"O}zdem, Ula{\c{s}}},
    title = "{Magnetic moments of pentaquark states in light-cone sum rules}",
    doi = "10.1140/epja/s10050-022-00700-2",
    journal = "Eur. Phys. J. A",
    volume = "58",
    number = "3",
    pages = "46",
    year = "2022"
}

@article{Wang:2022nqs,
    author = "Wang, Fu-Lai and Luo, Si-Qiang and Zhou, Hong-Yan and Liu, Zhan-Wei and Liu, Xiang",
    title = "{Exploring the electromagnetic properties of the {\ensuremath{\Xi}}c(',*)D{\textasciimacron}s* and {\ensuremath{\Omega}}c(*)D{\textasciimacron}s* molecular states}",
    eprint = "2210.02809",
    archivePrefix = "arXiv",
    primaryClass = "hep-ph",
    doi = "10.1103/PhysRevD.108.034006",
    journal = "Phys. Rev. D",
    volume = "108",
    number = "3",
    pages = "034006",
    year = "2023"
}

@article{Wang:2022tib,
    author = "Wang, Fu-Lai and Zhou, Hong-Yan and Liu, Zhan-Wei and Liu, Xiang",
    title = "{What can we learn from the electromagnetic properties of hidden-charm molecular pentaquarks with single strangeness?}",
    eprint = "2208.10756",
    archivePrefix = "arXiv",
    primaryClass = "hep-ph",
    doi = "10.1103/PhysRevD.106.054020",
    journal = "Phys. Rev. D",
    volume = "106",
    number = "5",
    pages = "054020",
    year = "2022"
}

@article{Ozdem:2024jty,
    author = {{\"O}zdem, Ula{\c{s}}},
    title = "{Analysis of the isospin eigenstate $\bar{D} \Sigma _c$, $\bar{D}^{*} \Sigma _c$, and $\bar{D} \Sigma _c^{*}$ pentaquarks by their electromagnetic properties}",
    eprint = "2401.12678",
    archivePrefix = "arXiv",
    primaryClass = "hep-ph",
    doi = "10.1140/epjc/s10052-024-13124-7",
    journal = "Eur. Phys. J. C",
    volume = "84",
    number = "8",
    pages = "769",
    year = "2024"
}

@article{Li:2024wxr,
    author = "Li, Hao-Song and Guo, Fei and Lei, Ya-Ding and Gao, Feng",
    title = "{Magnetic moments and axial charges of the octet hidden-charm molecular pentaquark family}",
    eprint = "2401.14767",
    archivePrefix = "arXiv",
    primaryClass = "hep-ph",
    doi = "10.1103/PhysRevD.109.094027",
    journal = "Phys. Rev. D",
    volume = "109",
    number = "9",
    pages = "094027",
    year = "2024"
}

@article{Li:2024jlq,
    author = "Li, Hao-Song",
    title = "{Molecular pentaquark magnetic moments in heavy pentaquark chiral perturbation theory}",
    eprint = "2401.14759",
    archivePrefix = "arXiv",
    primaryClass = "hep-ph",
    doi = "10.1103/PhysRevD.109.114039",
    journal = "Phys. Rev. D",
    volume = "109",
    number = "11",
    pages = "114039",
    year = "2024"
}

@article{Ozdem:2024yel,
    author = {{\"O}zdem, Ula{\c{s}}},
    title = "{Investigation on the electromagnetic properties of the $ D^{(*)} \Sigma _c^{(*)}$ molecules}",
    eprint = "2405.07273",
    archivePrefix = "arXiv",
    primaryClass = "hep-ph",
    doi = "10.1140/epja/s10050-024-01477-2",
    journal = "Eur. Phys. J. A",
    volume = "61",
    number = "1",
    pages = "10",
    year = "2025"
}

@article{Ozdem:2024rqx,
    author = {{\"O}zdem, Ula{\c{s}}},
    title = "{Elucidating the nature of hidden-charm pentaquark states with spin-32 through their electromagnetic form factors}",
    eprint = "2402.03802",
    archivePrefix = "arXiv",
    primaryClass = "hep-ph",
    doi = "10.1016/j.physletb.2024.138551",
    journal = "Phys. Lett. B",
    volume = "851",
    pages = "138551",
    year = "2024"
}

@article{Mutuk:2024ltc,
    author = "Mutuk, Halil and Kang, Xian-Wei",
    title = "{Unveiling the structure of hidden-bottom strange pentaquarks via magnetic moments}",
    eprint = "2405.07066",
    archivePrefix = "arXiv",
    primaryClass = "hep-ph",
    doi = "10.1016/j.physletb.2024.138772",
    journal = "Phys. Lett. B",
    volume = "855",
    pages = "138772",
    year = "2024"
}

@article{Mutuk:2024jxf,
    author = "Mutuk, Halil",
    title = "{Magnetic moments of hidden-bottom pentaquark states}",
    eprint = "2403.16616",
    archivePrefix = "arXiv",
    primaryClass = "hep-ph",
    doi = "10.1140/epjc/s10052-024-13263-x",
    journal = "Eur. Phys. J. C",
    volume = "84",
    number = "8",
    pages = "874",
    year = "2024"
}

@article{Ozdem:2024usw,
    author = {{\"O}zdem, Ula{\c{s}}},
    title = "{Insight into the nature of the $P_{c}(4457)$ and related pentaquarks}",
    eprint = "2409.09449",
    archivePrefix = "arXiv",
    primaryClass = "hep-ph",
    doi = "10.1140/epjc/s10052-025-14323-6",
    journal = "Eur. Phys. J. C",
    volume = "85",
    number = "6",
    pages = "624",
    year = "2025"
}

@article{Pascalutsa:2004je,
    author = "Pascalutsa, Vladimir and Vanderhaeghen, Marc",
    title = "{Magnetic moment of the Delta(1232)-resonance in chiral effective field theory}",
    eprint = "nucl-th/0412113",
    archivePrefix = "arXiv",
    reportNumber = "WM-04-124, JLAB-THY-05-292",
    doi = "10.1103/PhysRevLett.94.102003",
    journal = "Phys. Rev. Lett.",
    volume = "94",
    pages = "102003",
    year = "2005"
}

@article{Pascalutsa:2005vq,
    author = "Pascalutsa, Vladimir and Vanderhaeghen, Marc",
    title = "{Chiral effective-field theory in the Delta(1232) region: I. Pion electroproduction on the nucleon}",
    eprint = "hep-ph/0512244",
    archivePrefix = "arXiv",
    reportNumber = "WM-04-125, JLAB-THY-06-458",
    doi = "10.1103/PhysRevD.73.034003",
    journal = "Phys. Rev. D",
    volume = "73",
    pages = "034003",
    year = "2006"
}

@article{Pascalutsa:2007wb,
    author = "Pascalutsa, Vladimir and Vanderhaeghen, Marc",
    title = "{Chiral effective-field theory in the Delta(1232) region. II. Radiative pion photoproduction}",
    eprint = "0709.4583",
    archivePrefix = "arXiv",
    primaryClass = "hep-ph",
    reportNumber = "ECT*-07-19, WM-07-108, JLAB-THY-07-739",
    doi = "10.1103/PhysRevD.77.014027",
    journal = "Phys. Rev. D",
    volume = "77",
    pages = "014027",
    year = "2008"
}

@article{Can:2013zpa,
    author = "Can, K. U. and Erkol, G. and Isildak, B. and Oka, M. and Takahashi, T. T.",
    title = "{Electromagnetic properties of doubly charmed baryons in Lattice QCD}",
    eprint = "1306.0731",
    archivePrefix = "arXiv",
    primaryClass = "hep-lat",
    doi = "10.1016/j.physletb.2013.09.024",
    journal = "Phys. Lett. B",
    volume = "726",
    pages = "703--709",
    year = "2013"
}

@article{Can:2013tna,
    author = "Can, K. U. and Erkol, G. and Isildak, B. and Oka, M. and Takahashi, T. T.",
    title = "{Electromagnetic structure of charmed baryons in Lattice QCD}",
    eprint = "1310.5915",
    archivePrefix = "arXiv",
    primaryClass = "hep-lat",
    doi = "10.1007/JHEP05(2014)125",
    journal = "JHEP",
    volume = "05",
    pages = "125",
    year = "2014"
}

@article{Chernyak:1990ag,
    author = "Chernyak, V. L. and Zhitnitsky, I. R.",
    title = "{B meson exclusive decays into baryons}",
    doi = "10.1016/0550-3213(90)90612-H",
    journal = "Nucl. Phys. B",
    volume = "345",
    pages = "137--172",
    year = "1990"
}

@article{Braun:1988qv,
    author = "Braun, Vladimir M. and Filyanov, I. E.",
    title = "{QCD Sum Rules in Exclusive Kinematics and Pion Wave Function}",
    reportNumber = "LENINGRAD-88-1446",
    doi = "10.1007/BF01548594",
    journal = "Z. Phys. C",
    volume = "44",
    pages = "157",
    year = "1989"
}

@article{Balitsky:1989ry,
    author = "Balitsky, I. I. and Braun, Vladimir M. and Kolesnichenko, A. V.",
    title = "{Radiative Decay Sigma+ ---{\ensuremath{>}} p gamma in Quantum Chromodynamics}",
    doi = "10.1016/0550-3213(89)90570-1",
    journal = "Nucl. Phys. B",
    volume = "312",
    pages = "509--550",
    year = "1989"
}

@article{Ball:2002ps,
    author = "Ball, Patricia and Braun, V. M. and Kivel, N.",
    title = "{Photon distribution amplitudes in QCD}",
    eprint = "hep-ph/0207307",
    archivePrefix = "arXiv",
    reportNumber = "IPPP-02-40, DCPT-02-80",
    doi = "10.1016/S0550-3213(02)01017-9",
    journal = "Nucl. Phys. B",
    volume = "649",
    pages = "263--296",
    year = "2003"
}

@article{Wang:2010sh,
    author = "Wang, Zhi-Gang",
    title = "{Analysis of the scalar and axial-vector heavy diquark states with QCD sum rules}",
    eprint = "1008.4449",
    archivePrefix = "arXiv",
    primaryClass = "hep-ph",
    doi = "10.1140/epjc/s10052-010-1524-y",
    journal = "Eur. Phys. J. C",
    volume = "71",
    pages = "1524",
    year = "2011"
}

@article{Kleiv:2013dta,
    author = "Kleiv, R. T. and Steele, T. G. and Zhang, Ailin and Blokland, Ian",
    title = "{Heavy-light diquark masses from QCD sum rules and constituent diquark models of tetraquarks}",
    eprint = "1304.7816",
    archivePrefix = "arXiv",
    primaryClass = "hep-ph",
    doi = "10.1103/PhysRevD.87.125018",
    journal = "Phys. Rev. D",
    volume = "87",
    number = "12",
    pages = "125018",
    year = "2013"
}

@article{Nozawa:1990gt,
    author = "Nozawa, S. and Leinweber, D. B.",
    title = "{Electromagnetic form-factors of spin 3/2 baryons}",
    reportNumber = "TRI-PP-90-19",
    doi = "10.1103/PhysRevD.42.3567",
    journal = "Phys. Rev. D",
    volume = "42",
    pages = "3567--3571",
    year = "1990"
}

@article{Pascalutsa:2006up,
    author = "Pascalutsa, Vladimir and Vanderhaeghen, Marc and Yang, Shin Nan",
    title = "{Electromagnetic excitation of the Delta(1232)-resonance}",
    eprint = "hep-ph/0609004",
    archivePrefix = "arXiv",
    reportNumber = "JLAB-THY-06-537",
    doi = "10.1016/j.physrep.2006.09.006",
    journal = "Phys. Rept.",
    volume = "437",
    pages = "125--232",
    year = "2007"
}

@article{Ramalho:2009vc,
    author = "Ramalho, G. and Pena, M. T. and Gross, Franz",
    title = "{Electric quadrupole and magnetic octupole moments of the Delta}",
    eprint = "0902.4212",
    archivePrefix = "arXiv",
    primaryClass = "hep-ph",
    reportNumber = "JLAB-THY-09-951",
    doi = "10.1016/j.physletb.2009.06.052",
    journal = "Phys. Lett. B",
    volume = "678",
    pages = "355--358",
    year = "2009"
}

@article{Balitsky:1987bk,
    author = "Balitsky, I. I. and Braun, Vladimir M.",
    title = "{Evolution Equations for QCD String Operators}",
    reportNumber = "LENINGRAD-87-1351",
    doi = "10.1016/0550-3213(89)90168-5",
    journal = "Nucl. Phys. B",
    volume = "311",
    pages = "541--584",
    year = "1989"
}

@article{Belyaev:1985wza,
    author = "Belyaev, V. M. and Blok, B. Yu.",
    title = "{CHARMED BARYONS IN QUANTUM CHROMODYNAMICS}",
    doi = "10.1007/BF01560689",
    journal = "Z. Phys. C",
    volume = "30",
    pages = "151",
    year = "1986"
}

@article{ParticleDataGroup:2024cfk,
    author = "Navas, S. and others",
    collaboration = "Particle Data Group",
    title = "{Review of particle physics}",
    doi = "10.1103/PhysRevD.110.030001",
    journal = "Phys. Rev. D",
    volume = "110",
    number = "3",
    pages = "030001",
    year = "2024"
}

@article{Ioffe:2005ym,
    author = "Ioffe, B. L.",
    title = "{QCD at low energies}",
    eprint = "hep-ph/0502148",
    archivePrefix = "arXiv",
    doi = "10.1016/j.ppnp.2005.05.001",
    journal = "Prog. Part. Nucl. Phys.",
    volume = "56",
    pages = "232--277",
    year = "2006"
}

@article{Narison:2018nbv,
    author = "Narison, Stephan",
    editor = "Narison, St{\'e}phan",
    title = "{$\overline{\rm m}_{c,b,}<\alpha_sG^2>$ and $\alpha_s$ from Heavy Quarkonia}",
    doi = "10.1016/j.nuclphysbps.2018.12.026",
    journal = "Nucl. Part. Phys. Proc.",
    volume = "300-302",
    pages = "153--164",
    year = "2018"
}

@article{Rohrwild:2007yt,
    author = "Rohrwild, J.",
    title = "{Determination of the magnetic susceptibility of the quark condensate using radiative heavy meson decays}",
    eprint = "0708.1405",
    archivePrefix = "arXiv",
    primaryClass = "hep-ph",
    doi = "10.1088/1126-6708/2007/09/073",
    journal = "JHEP",
    volume = "09",
    pages = "073",
    year = "2007"
}

@article{Ozdem:2024txt,
    author = {{\"O}zdem, Ula{\c{s}}},
    title = "{Investigating the underlying structure of vector hidden-charm tetraquark states via their electromagnetic characteristics}",
    eprint = "2412.06447",
    archivePrefix = "arXiv",
    primaryClass = "hep-ph",
    doi = "10.1103/PhysRevD.111.054009",
    journal = "Phys. Rev. D",
    volume = "111",
    number = "5",
    pages = "054009",
    year = "2025"
}

@article{Ozdem:2024dbq,
    author = {{\"O}zdem, Ula{\c{s}}},
    title = "{Unveiling the underlying structure of axial-vector bottom-charm tetraquarks in the light of their magnetic moments}",
    eprint = "2403.16191",
    archivePrefix = "arXiv",
    primaryClass = "hep-ph",
    doi = "10.1007/JHEP05(2024)301",
    journal = "JHEP",
    volume = "05",
    pages = "301",
    year = "2024"
}

@article{Azizi:2023gzv,
    author = {Azizi, K. and {\"O}zdem, U.},
    title = "{Exploring the magnetic dipole moments of $ {T}_{QQ\overline{q}\overline{s}} $ and $ {T}_{QQ\overline{s}\overline{s}} $ states in the framework of QCD light-cone sum rules}",
    eprint = "2301.07713",
    archivePrefix = "arXiv",
    primaryClass = "hep-ph",
    doi = "10.1007/JHEP03(2023)166",
    journal = "JHEP",
    volume = "03",
    pages = "166",
    year = "2023"
}

@article{Ozdem:2025fks,
    author = {{\"O}zdem, Ula{\c{s}}},
    title = "{Probing the electromagnetic structure of the $P_c(4337)^+$ pentaquark: insights from a diquark{\textendash}diquark{\textendash}antiquark picture for $J^P = \frac{1}{2}^-$ and $\frac{3}{2}^-$ states}",
    eprint = "2506.04345",
    archivePrefix = "arXiv",
    primaryClass = "hep-ph",
    doi = "10.1140/epjc/s10052-025-14439-9",
    journal = "Eur. Phys. J. C",
    volume = "85",
    number = "6",
    pages = "704",
    year = "2025"
}

@article{Ozdem:2024rch,
    author = {{\"O}zdem, Ula{\c{s}}},
    title = "{Shedding light on the nature of the Pcs(4459) pentaquark state}",
    eprint = "2411.11442",
    archivePrefix = "arXiv",
    primaryClass = "hep-ph",
    doi = "10.1103/PhysRevD.111.074038",
    journal = "Phys. Rev. D",
    volume = "111",
    number = "7",
    pages = "074038",
    year = "2025"
}

@article{Ozdem:2025ncd,
    author = {{\"O}zdem, Ula{\c{s}}},
    title = "{Unveiling the electromagnetic structure and intrinsic dynamics of spin-3/2 hidden-charm pentaquarks: A comprehensive QCD analysis}",
    eprint = "2504.13488",
    archivePrefix = "arXiv",
    primaryClass = "hep-ph",
    doi = "10.1088/1674-1137/ade95a",
    journal = "Chin. Phys.",
    volume = "49",
    number = "10",
    pages = "103106",
    year = "2025"
}

@article{Vujmilovic:2025czt,
    author = "Vujmilovic, Ivan and Collins, Sara and Leskovec, Luka and Prelovsek, Sasa",
    title = "{Electromagnetic form factors and structure of the $T_{bb}$ tetraquark from lattice QCD}",
    eprint = "2510.17549",
    archivePrefix = "arXiv",
    primaryClass = "hep-lat",
    month = "10",
    year = "2025"
}

@article{Clymton:2025hez,
    author = "Clymton, Samson and Kim, Hyun-Chul and Mart, Terry",
    title = "{Production mechanism of hidden-charm pentaquark states Pcc{\textasciimacron}s with strangeness S=-1}",
    eprint = "2504.07693",
    archivePrefix = "arXiv",
    primaryClass = "hep-ph",
    reportNumber = "INHA-NTG-03/2025",
    doi = "10.1103/2wvg-jlxp",
    journal = "Phys. Rev. D",
    volume = "112",
    number = "1",
    pages = "014041",
    year = "2025"
}

@article{Zhu:2025abk,
    author = "Zhu, Sheng-He and Wang, Fu-Lai and Liu, Xiang",
    title = "{Electromagnetic characteristics as probes into the inner structures of the predicted $\Xi_c^{(',*)}D^{(*)}_s$ molecular states}",
    eprint = "2510.18492",
    archivePrefix = "arXiv",
    primaryClass = "hep-ph",
    month = "10",
    year = "2025"
}

@article{Ozdem:2025ion,
    author = {{\"O}zdem, Ula{\c{s}}},
    title = "{Electromagnetic form factors: a window into the $D\Lambda _c$, $D^*\Lambda _c$, and $D\Lambda _c^*$ molecular structure}",
    eprint = "2511.16052",
    archivePrefix = "arXiv",
    primaryClass = "hep-ph",
    doi = "10.1140/epjc/s10052-026-15940-5",
    journal = "Eur. Phys. J. C",
    volume = "86",
    number = "6",
    pages = "675",
    year = "2026"
}

@article{Ozdem:2025jda,
    author = {{\"O}zdem, Ula{\c{s}}},
    title = "{Electromagnetic tomography of spin-$ \frac{3}{2} $ hidden-charm strange pentaquarks}",
    eprint = "2510.26893",
    archivePrefix = "arXiv",
    primaryClass = "hep-ph",
    doi = "10.1007/JHEP02(2026)207",
    journal = "JHEP",
    volume = "02",
    pages = "207",
    year = "2026"
}

@article{Leinweber:1990dv,
    author = "Leinweber, Derek B. and Woloshyn, R. M. and Draper, Terrence",
    title = "{Electromagnetic structure of octet baryons}",
    reportNumber = "TRI-PP-90-52, UK-PP-90-09",
    doi = "10.1103/PhysRevD.43.1659",
    journal = "Phys. Rev. D",
    volume = "43",
    pages = "1659--1678",
    year = "1991"
}

@article{Wang:2025pjt,
    author = "Wang, Zhi-Gang and Liu, Yang",
    title = "{Analysis of the hidden-charm pentaquark candidates in the $J/\psi \Xi$ mass spectrum via the QCD sum rules}",
    eprint = "2511.13067",
    archivePrefix = "arXiv",
    primaryClass = "hep-ph",
    doi = "10.1016/j.nuclphysb.2026.117456",
    journal = "Nucl. Phys. B",
    volume = "1027",
    pages = "117456",
    year = "2026"
}

@article{Ozdem:2024lpk,
    author = {{\"O}zdem, U. and Azizi, K.},
    title = "{Electromagnetic properties of vector doubly charmed tetraquark states}",
    eprint = "2401.04798",
    archivePrefix = "arXiv",
    primaryClass = "hep-ph",
    doi = "10.1103/PhysRevD.109.114019",
    journal = "Phys. Rev. D",
    volume = "109",
    number = "11",
    pages = "114019",
    year = "2024"
}

@article{Mutuk:2024ach,
    author = "Mutuk, Halil",
    title = "{Magnetic moments of hidden-charm pentaquarks in the diquark{\textendash}diquark{\textendash}antiquark scheme}",
    eprint = "2411.16486",
    archivePrefix = "arXiv",
    primaryClass = "hep-ph",
    doi = "10.1016/j.cjph.2025.07.030",
    journal = "Chin. J. Phys.",
    volume = "97",
    pages = "1406--1414",
    year = "2025"
}

@article{Duan:2024uuf,
    author = "Duan, Feng-Bo and Wang, Qi-Nan and Yang, Zi-Yan and Chen, Xu-Liang and Chen, Wei",
    title = "{Doubly charmed pentaquark states in QCD sum rules}",
    eprint = "2401.10078",
    archivePrefix = "arXiv",
    primaryClass = "hep-ph",
    doi = "10.1103/PhysRevD.109.094018",
    journal = "Phys. Rev. D",
    volume = "109",
    number = "9",
    pages = "094018",
    year = "2024"
}

@article{Pimikov:2019dyr,
    author = "Pimikov, Alexandr and Lee, Hee-Jung and Zhang, Pengming",
    title = "{Hidden charm pentaquarks with color-octet substructure in QCD Sum Rules}",
    eprint = "1908.04459",
    archivePrefix = "arXiv",
    primaryClass = "hep-ph",
    doi = "10.1103/PhysRevD.101.014002",
    journal = "Phys. Rev. D",
    volume = "101",
    number = "1",
    pages = "014002",
    year = "2020"
}

@article{Zhu:2015bba,
    author = "Zhu, Ruilin and Qiao, Cong-Feng",
    title = "{Pentaquark states in a diquark{\textendash}triquark model}",
    eprint = "1510.08693",
    archivePrefix = "arXiv",
    primaryClass = "hep-ph",
    doi = "10.1016/j.physletb.2016.03.022",
    journal = "Phys. Lett. B",
    volume = "756",
    pages = "259--264",
    year = "2016"
}

@article{Lebed:2015tna,
    author = "Lebed, Richard F.",
    title = "{The Pentaquark Candidates in the Dynamical Diquark Picture}",
    eprint = "1507.05867",
    archivePrefix = "arXiv",
    primaryClass = "hep-ph",
    doi = "10.1016/j.physletb.2015.08.032",
    journal = "Phys. Lett. B",
    volume = "749",
    pages = "454--457",
    year = "2015"
}

@article{Maiani:2015vwa,
    author = "Maiani, L. and Polosa, A. D. and Riquer, V.",
    title = "{The New Pentaquarks in the Diquark Model}",
    eprint = "1507.04980",
    archivePrefix = "arXiv",
    primaryClass = "hep-ph",
    doi = "10.1016/j.physletb.2015.08.008",
    journal = "Phys. Lett. B",
    volume = "749",
    pages = "289--291",
    year = "2015"
}

@article{LHCb:2025lhk,
    author = "Aaij, R. and others",
    collaboration = "LHCb",
    title = "{Observation of the ${{{\varLambda } ^0_{b}} \!\rightarrow {{J \hspace{-1.66656pt}/\hspace{-1.111pt}\psi }} {{\varXi } ^-} {{K} ^+} }$ and ${{{{\varXi } ^0_{b}} \!\rightarrow {{J \hspace{-1.66656pt}/\hspace{-1.111pt}\psi }} {{\varXi } ^-} {{\pi } ^+} }}$ decays}",
    eprint = "2501.12779",
    archivePrefix = "arXiv",
    primaryClass = "hep-ex",
    reportNumber = "CERN-EP-2024-337 LHCb-PAPER-2024-049, CERN-EP-2024-337, LHCb-PAPER-2024-049",
    doi = "10.1140/epjc/s10052-025-14129-6",
    journal = "Eur. Phys. J. C",
    volume = "85",
    number = "7",
    pages = "812",
    year = "2025"
}

@article{CMS:2024vnm,
    author = "Hayrapetyan, Aram and others",
    collaboration = "CMS",
    title = "{Observation of the $\Lambda_\text{b}^0\to J/\psi\Xi^-K^+$ decay}",
    eprint = "2401.16303",
    archivePrefix = "arXiv",
    primaryClass = "hep-ex",
    reportNumber = "CMS-BPH-22-002, CERN-EP-2024-006",
    doi = "10.1140/epjc/s10052-024-13114-9",
    journal = "Eur. Phys. J. C",
    volume = "84",
    number = "10",
    pages = "1062",
    year = "2024"
}

@article{Oset:2024fbk,
    author = "Oset, E. and Roca, L. and Whitehead, M.",
    title = "{Production of pentaquarks with hidden charm and double strangeness in {\ensuremath{\Xi}}b and {\ensuremath{\Omega}}b decays}",
    eprint = "2406.16504",
    archivePrefix = "arXiv",
    primaryClass = "hep-ph",
    doi = "10.1103/PhysRevD.110.034016",
    journal = "Phys. Rev. D",
    volume = "110",
    number = "3",
    pages = "034016",
    year = "2024"
}

@article{Roca:2025zyi,
    author = "Roca, L. and Song, J. and Oset, E.",
    title = "{Study of hidden-charm, doubly-strange pentaquarks in $\Lambda _b\rightarrow J/\psi \Xi ^- K^+$ and $\Xi _b\rightarrow J/\psi \Xi ^- \pi ^+$}",
    eprint = "2509.19840",
    archivePrefix = "arXiv",
    primaryClass = "hep-ph",
    doi = "10.1140/epjc/s10052-025-15280-w",
    journal = "Eur. Phys. J. C",
    volume = "86",
    number = "2",
    pages = "100",
    year = "2026"
}
\end{document}